\documentclass[a4paper,fleqn,usenatbib]{mnras}

\usepackage{newtxtext,newtxmath}
\usepackage[T1]{fontenc}
\usepackage{ae,aecompl}
\usepackage{graphicx}	
\usepackage{longtable}

\title[Outlier galaxies in HST]{Automatic identification of outliers in Hubble Space Telescope galaxy images}

\author[Lior Shamir]{
Lior Shamir,$^{1}$\thanks{E-mail: lshamir@mtu.edu}
\\
$^{1}$Kansas State University, Manhattan, KS 65506, USA
}

\date{Accepted XXX. Received YYY; in original form ZZZ}

\pubyear{2020}

\begin{document}
\label{firstpage}
\pagerange{\pageref{firstpage}--\pageref{lastpage}}

\maketitle

\begin{abstract}

Rare extragalactic objects can carry substantial information about the past, present, and future universe. Given the size of astronomical databases in the information era it can be assumed that very many outlier galaxies are included in existing and future astronomical databases. However, manual search for these objects is impractical due to the required labor, and therefore the ability to detect such objects largely depends on computer algorithms. This paper describes an unsupervised machine learning algorithm for automatic detection of outlier galaxy images, and its application to several Hubble Space Telescope fields. The algorithm does not require training, and therefore is not dependent on the preparation of clean training sets. The application of the algorithm to a large collection of galaxies detected a variety of outlier galaxy images. The algorithm is not perfect in the sense that not all objects detected by the algorithm are indeed considered outliers, but it reduces the dataset by two orders of magnitude to allow practical manual identification. The catalogue contains 147 objects that would be very difficult to identify without using automation.

\end{abstract}


\section{Introduction}
\label{introduction}

While most galaxies can be classified into known morphological types, some galaxies do not fit in any of these common morphologies, and are considered ``peculiar''. The ``peculiarity'' of a galaxy is normally determined by its visual appearance, and the classification of a galaxy as peculiar is not strictly defined \citep{nairn1997peculiar}. 
However, these galaxies can carry important information about galaxy evolution \citep{gillman2020peculiar}, and are therefore of scientific importance \citep{bettoni2001gas,casasola2004gas,abraham2001morphological}.

One of the first notable attempts to profile peculiar galaxies was the Atlas of Peculiar Galaxies \citep{arp1966atlas,arp1975catalogue}, that was prepared manually. Other notable efforts to prepare catalogues of peculiar galaxies include the catalog of collisional ring galaxies \citep{madore2009atlas}. Digital sky surveys provided very large datasets of galaxies, making the identification of peculiar galaxies more efficient. For instance, \cite{kaviraj2010peculiar} used a set of 70 early-type peculiar systems in Sloan Digital Sky Survey (SDSS) stripe 82. Another example is the catalogue of \citep{nair2010catalog}, providing information about the morphology of $\sim1.4\cdot10^4$ galaxies. During the preparation of the catalogue, numerous peculiar galaxies were identified. \cite{taylor2005ubvr} compiled a collection of 142 galaxies that included spiral, irregular, and interacting galaxies by using the Vatican Advanced Technology Telescope. But because the analysis was performed manually it was limited by the number of galaxies that were analyzed \citep{nair2010catalog}.  

Because manual analysis is naturally slow, it does not allow to handle very large databases of galaxies, or requires very substantial efforts. For instance, the catalogue of \citep{arp1966atlas} took about 14 years to complete. In attempt to increase the throughput of the detection of peculiar galaxies, crowdsourcing was used by allowing volunteers to annotate galaxies, leading to the identification of ``Hanny's Voorwerp" \citep{lintott2009galaxy}. That approach also led to the identification of a high number of ring galaxies \citep{finkelman2012polar,buta2017galactic}.

Hubble Space Telescope (HST) was able to provide deeper and more detailed images of galaxies, providing much more detailed images of objects that cannot be analyzed morphologically by Earth-based sky surveys such as SDSS and the Panoramic Survey Telescope and Rapid Response System (Pan-STARRS). Therefore, HST allows to identify peculiar galaxies in much higher redshifts compared to Earth-based surveys. Although HST surveys are smaller than Earth-based surveys such as SDSS, surveys such as the Cosmic Evolution Survey (COSMOS) still contain more than $2\cdot10^6$ galaxies \citep{scoville2007cosmic}.

While current sky surveys such as SDSS, Pan-STARRS, and the Dark Energy Survey (DES) are already far too large to allow comprehensive manual analysis, future digital sky surveys such as the Vera Rubin Observatory will acquire even more data and a far higher number of celestial objects. To allow using these data effectively, methods based on computer analysis of galaxy images have been proposed. These include model-driven methods such as GALFIT \citep{pen02}, GIM2D \citep{sim99}, CAS \citep{con03}, Gini \citep{abraham2003new}, Ganalyzer \citep{sha11}, and SpArcFiRe \citep{davis2014sparcfire}, and methods based on machine learning \citep{sha09,huertas2009robust,banerji2010,kum14,dieleman2015rotation,graham2019galaxy,mittal2019data,hosny2020classification,cecotti2020rotation,cheng2020optimizing}. The application of these methods led to catalogues \citep{huertas2015catalog,huertas2015morphologies,shamir2014automatic,kuminski2016computer,goddard2020}. Machine learning algorithms were also used to identify unusual galaxies, such as galaxy mergers \citep{margalef2020detecting}, showing that galaxy mergers can be identified automatically even when training a machine learning system with just regular isolated galaxies.

Model-driven approaches were used in the past to detect specific types of galaxies such as ring galaxies \citep{timmis2017catalog,shamir2020automatic} or gravitational lenses \citep{jacobs2019finding}. Comparing these algorithms to datasets prepared manually showed that computers were not able to achieve the same level of completeness of manual detection, but can compensate for that weakness by their ability to scan much larger datasets \citep{shamir2020automatic}. The main weakness of model-driven algorithms is that they can be developed only when the morphology of interest is known, and therefore cannot detect unknown objects of types that have not been observed before.

Machine learning is often applied by training a system from the data rather than tailoring a specific algorithm. The majority of machine learning methods proposed for galaxy image analysis are based on supervised machine learning, in which a machine learning system is trained with annotated ``ground truth'' to classify new unseen samples. Such supervised machine learning systems might not be effective for detecting outlier galaxy images that have not been seen before, as no samples are available to train such systems. To be able to identify peculiar galaxies automatically, a machine learning system needs to be able to identify forms of galaxies that are not present in the dataset with which the system was trained. Therefore, for the identification of such outlier galaxies, unsupervised machine learning is required. Additionally, it needs to be able to filter false positives effectively, as due to the large number of objects even a small false positive rate would lead to a very high number of false positives, making such system impractical.

\section{Method}
\label{method}

Non-parametric approaches such as deep convolutional neural networks (DCNNs) have been adjusted to the task of outlier image detection. One of the common approaches to outlier image detection using deep neural networks is by using auto-encoders, such that outliers can be detected by the reconstruction loss \citep{amarbayasgalan2018unsupervised,chen2018evolutionary}, and were also applied to outlier galaxy detection \citep{venkat2020}. Deep neural networks have shown promising performance for the task of identifying merging systems in datasets of isolated galaxies \citep{margalef2020detecting}. Since in a universe of isolated galaxies a merging system would be considered an outlier, the performance of the algorithm is an indication of the ability to detect outlier galaxies. 

While deep neural networks provide promising performance in detecting outlier galaxies, they also require large clean training sets, and their ``black box'' nature makes them more difficult to identify specific elements that make certain galaxies marked as outliers. The purpose of this work is to use algorithms that do not require labeling, so that galaxies of types that are not known can also be detected. To perform unsupervised machine learning of galaxy images, each galaxy image is converted into a comprehensive set of numerical image content descriptors that reflect the visual content of the image. That is, each image is represented by a vector of numbers that correspond to the visual content. The set of numerical image content descriptors \citep{shamir2008wndchrm} has been shown efficacy in analysis of galaxy images \citep{sha09,kum14,kuminski2016computer}, including certain tasks in unsupervised analysis of galaxy images \citep{shamir2012automatic,shamir2013automatic,schutter2015galaxy}.

In summary, the set of numerical image content descriptors include edge statistics, Radon transform \citep{lim1990two}, texture descriptors such Tamura textures \citep{tamura1978textural}, Haralick textures \citep{haralick1973textural}, and Gabor textures \citep{fogel1989gabor}, distribution of pixel intensities multi-scale histograms \citep{hadjidemetriou2001spatial}, Zernike polynomials \citep{teague1980image}, the Gini coefficient \citep{abraham2003new}, image entropy, Chebyshev statistics, and box-counting fractals \citep{wu1992texture}. These numerical image content descriptors are described in detail in \citep{shamir2008wndchrm,shamir2010impressionism,shamir2013automatic,schutter2015galaxy,shamir2016morphology}. 

To obtain more information from each galaxy image, the numerical image content descriptors are extracted from the raw pixels, but also from several image transforms. These include the Fourier transform, Chebyshev transform, Wavelet (symlet 5, level 1) transform, and combinations of these transforms \citep{shamir2008wndchrm,shamir2010impressionism}. The source code of the method is open and publicly available \citep{shamir2017udat}.

When using a high number of numerical content descriptors, it is expected that some of them would not reflect information relevant to the difference between regular and irregular galaxies. Since the algorithm aims at identifying also types of galaxies that have not been seen before, previously collected data cannot be used for that task. To rank and weight the content descriptors by the information they provide in identifying outlier galaxy images without using annotated samples, the entropy of each feature {\it f} is used as shown in Equation~\ref{entropy}.

\begin{equation}
W_f=-1\cdot \Sigma_i P_i \cdot \log P_i  ,
\label{entropy}
\end{equation}
where $P_i$ is the frequency of the values in the i{\it th} bin of a 10-bin histogram of the values of that feature. $W_f$ is the entropy of the feature, which is used as the weight. When the entropy of the feature is low, the feature values are more consistent, and that consistency can be used as an indication that the numerical content descriptor is informative for reflecting the morphology of the galaxies in the dataset.
 
The dissimilarity between each pair of galaxies can be computed by using the Earth Mover's Distance (EMD), which is an effective way of comparing vectors, and commonly using in machine learning tasks \citep{rubner2000earth,ruzon2001edge}. EMD can be conceptualized as an optimization problem in which the solution is the minimum work required to fill a set of holes in space with the mass of Earth, and the unit of work is the work required to move an Earth unit by a distance unit. Equation~\ref{emd} shows the EMD optimization problem.

\begin{equation}
\label{emd}
Work(X,Y,F)=\Sigma_{i=1}^n \Sigma_{j=1}^n f_{i,j}d_{i,j},
\end{equation}

where X and Y are the weighted feature vectors ${(Wx_1,x_1).....(Wx_n,x_n)}$ of size n, $f_{i,j}$ is the flow between $X_i$ and $Y_j$, and $W$ is the vector of weights determined for all features by Equation~\ref{entropy}. The flow F is the solution of the following linear programming problem: \newline \newline
$\Sigma_{i=1}^n \Sigma_{j=1}^n f_{i,j} = \min ( \Sigma_{i=1}^n Wx_i , \Sigma_{j=1}^n Wy_j ) $     \newline  \newline
With the following constraints: \newline \newline
$Wx_i \geq \Sigma_{j=1}^n f_{i,j}$      \newline \newline
$Wy_j \geq \Sigma_{i=1}^n f_{i,j}$     \newline \newline

The earth mover’s distance between X and Y is then defined as: \newline
$EMD(X,Y)=\frac{Work(X,Y,F)}{\Sigma_{i=1}^n \Sigma_{j=1}^n f_{i,j} } $

More details about the EMD vector comparison can be found in \citep{rubner2000earth,ruzon2001edge}. The EMD is used to measure the distance between the histograms of all sets of numerical image content descriptors described in \citep{shamir2008wndchrm,shamir2010impressionism}. The sum of all distances of all histograms determines the distance between the two galaxies. The distances measured between different pairs of galaxies can be compared to distances between other pairs of galaxies to provide an estimation of the level of similarity or difference between each pair of galaxies in a dataset.

Once the similarity between each pair of galaxy images can be measured, the outlier galaxy images can be detected. A simple way of identifying outlier galaxy images is by identifying the galaxy x such that $Max_{x} ( Min_{x,y} ( d(x,y)))$. That is, the galaxy image that is the most likely to be an outlier image is the galaxy that its distance to its most similar galaxy is the highest compared to all other galaxies. However, that criterion might lead to undetected outlier galaxies. When a dataset is large, it is possible that even a rare galaxy type will appear more than once in that dataset. For instance, the dataset of the Vera Rubin Observatory is expected to image $\sim10^{10}$ extra galactic objects, and therefore even a rare one-in-a-million object is expected to be present in that dataset $\sim10^4$ times. Therefore, even rare objects might have one or more objects in the datasets that is similar to them. That can lead to low maximum distance for these objects, and will lead to inability of the algorithm to identify outlier objects.

To avoid a situation in which a small number of outlier objects that are similar to each other are not detected, the distances of the objects from all other objects are sorted, and the R shortest distance is used as the minimum distance between the object and all other objects in the dataset. That means that if R-1 objects that are similar to the target object exist in the dataset, the distances between these objects and the target object will not affect the results. By using the rank R, a small number of objects that are similar to a certain object will not lead to inability to detect that object. The value of R should be determined based on the size of the dataset. The larger the dataset is, the more likely that a certain rare object will have other objects in the dataset that are similar to it. Therefore, a larger dataset will require a higher value of R to be able to detect outlier galaxies.

The R parameter is used by the algorithm to control the rank of the neighbor by which the distance of the sample from the dataset is measured. The value of R allows to reduce the impact of galaxies with elements that are less common in the dataset. Outlier detection algorithms might be dependent on the distribution of the samples in the dataset. For instance, if most galaxies in the dataset are small, larger galaxies might be identified as outliers. However, due to the R parameter, the distance between a sample and the rest of the dataset is determined by the distance between the sample and its R{\it th} closest neighbor. Therefore, in the case of uneven distribution of the size of the objects such that large objects are rare, the presence of more than R large objects in the dataset should theoretically prevent from large objects be identified as outlier due to their size alone. That it, if more than R large galaxies are present in the dataset, the R{\it th} neighbor of a large galaxy is expected to be a large galaxy, and therefore the distance that determines whether the sample is an outlier should not be large because the R{\it th} neighbor is small. Due to noise and the imperfectness of the distances it is expected that exactly R large galaxies might not be sufficient to avoid large objects identified as outliers, but in large datasets the number of large objects is expected to be much higher than the value of R, and therefore the R{\it th} nearest neighbor itself is expected to be a large object. That should ensure that even if large objects are the minority of the objects in the dataset, that should not lead to large objects being identified as outliers.

\section{Data}
\label{data}

\begin{figure}
\centering
\includegraphics[scale=0.6]{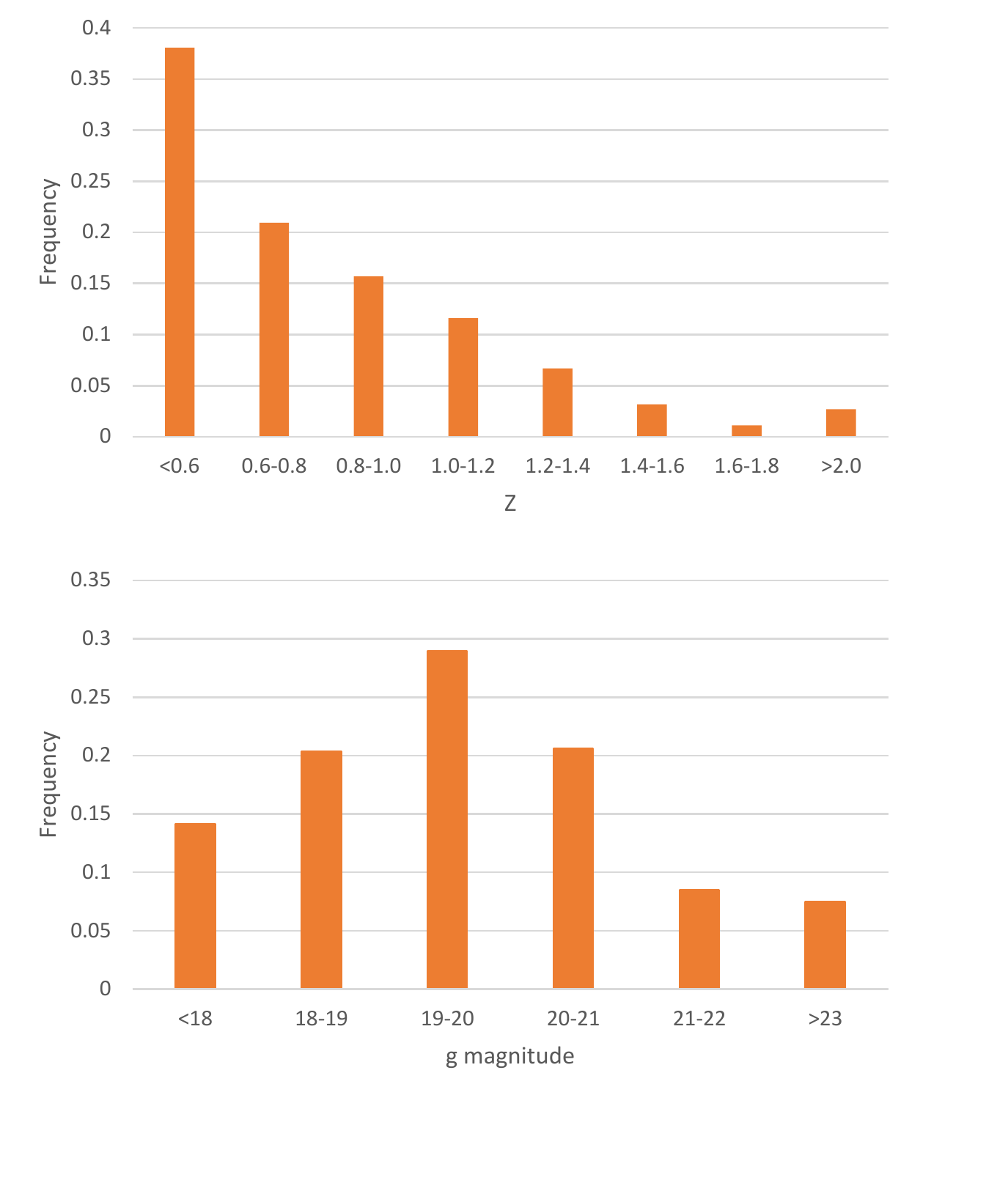}
\caption{The redshift and g magnitude distribution of the objects.}
\label{mag_z}
\end{figure}

The data is taken from several HST fields that make the Cosmic Assembly Near-infrared Deep Extragalactic Legacy Survey (CANDELS). CANDELS \citep{grogin2011candels,koekemoer2011candels} covers five fields, which are GOODS-N, GOODS-S, EGS, UDS, and COSMOS \citep{grogin2011candels}. Sources were detected by applying SExtractor \citep{bertin1996sextractor} on the F814W band and selecting sources with 4$\sigma$ or higher magnitude compared to the background. The sources were then separated by using the {\it Subimage} tool of Montage \citep{berriman2004montage}. The images were FITS images of dimensionality of 122$\times$122 pixels, and these images were converted to TIFF format for the image processing. The total number of objects was 176,808. The redshift and g magnitude distribution of these objects is shown in Figure~\ref{mag_z}.  


\section{Results}
\label{results}

The method described in Section~\ref{method} was applied to the data described in Section~\ref{data}. The method assigns each galaxy that it analyzes with a score of ``peculiarity'', and therefore allows to find the galaxies that are the most likely to be indeed peculiar. The 1,100 galaxy images with the highest likelihood to be peculiar according to the method were examined manually, making a selection of $\sim$0.5\% of the number of galaxies, but a sufficiently small number of galaxies to allow manual analysis.

\begin{figure}
\centering
\includegraphics[scale=0.75]{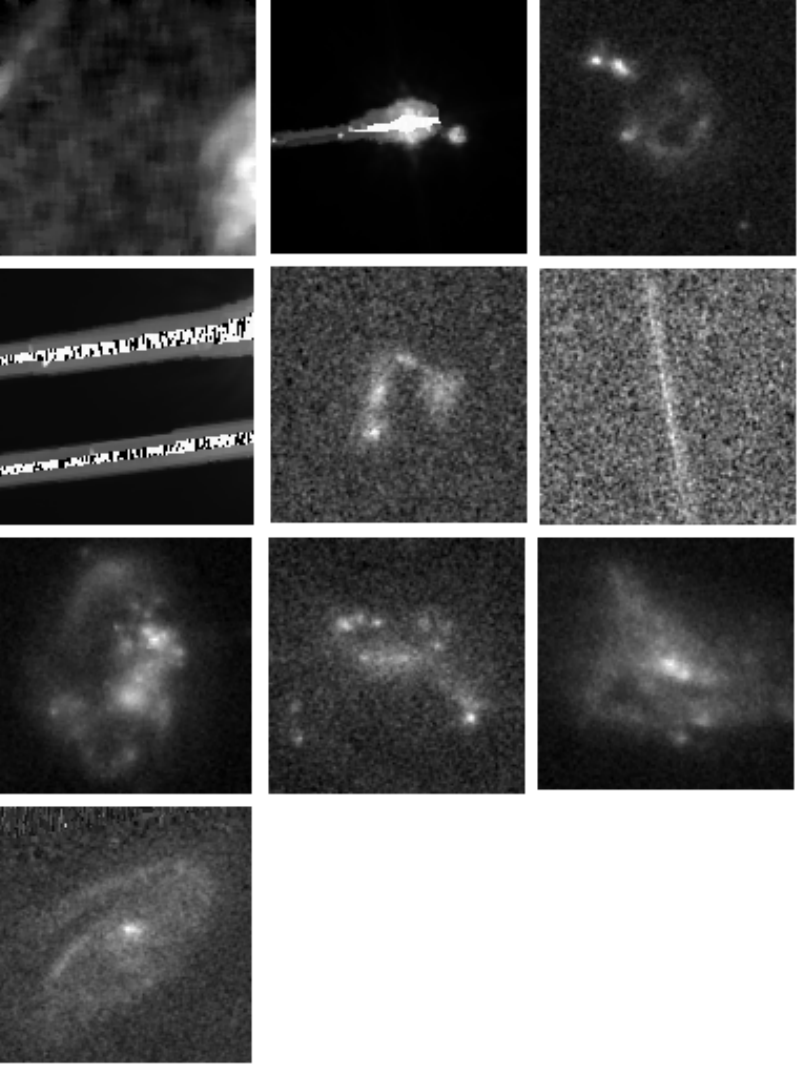}
\caption{The top 10 outlier galaxy images as ranked by the algorithm.}
\label{top10}
\end{figure}

\begin{figure}
\centering
\includegraphics[scale=0.75]{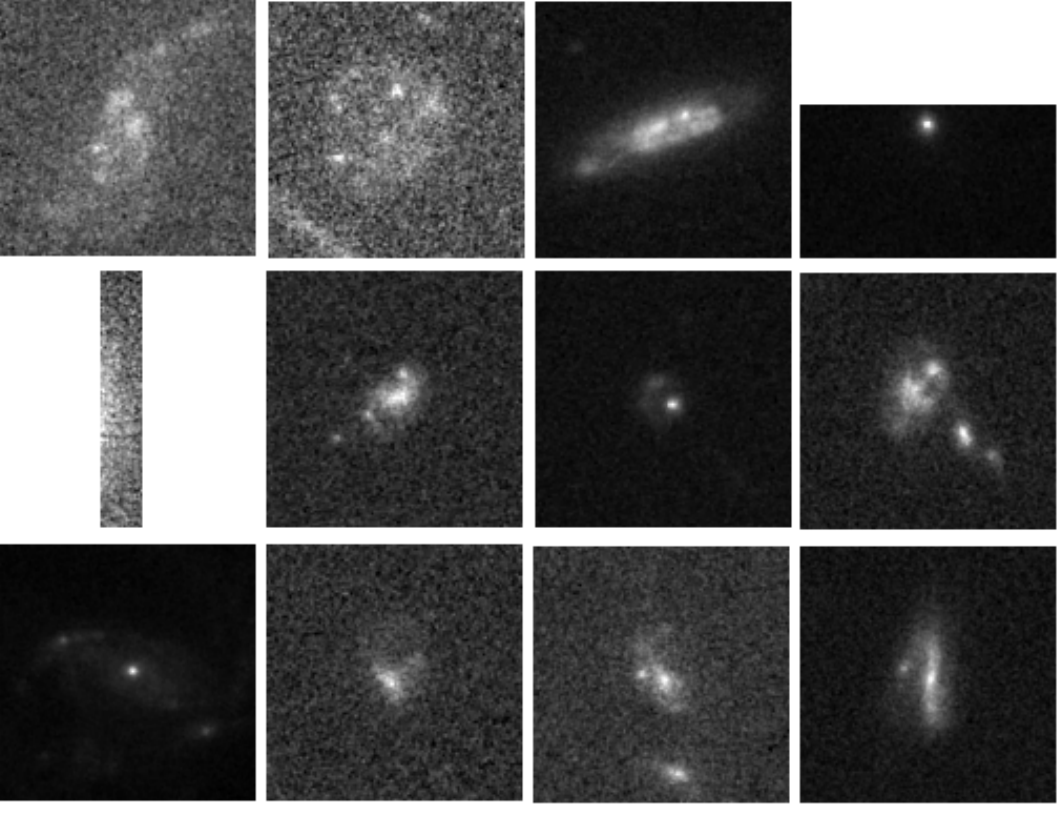}
\caption{Objects identified by the algorithm as outliers that are not indeed outlier galaxy images.}
\label{false_positives}
\end{figure}

Figure~\ref{top10} shows the top 10 galaxies identified as outlier galaxy images by the algorithm. As can be seen in the image, many of these galaxies are not considered peculiar, but are detected by the algorithm as galaxies that do not have a high number of similar images in the dataset. Due to the presence of a large number of objects that are not outlier galaxies, manual analysis is required for sorting the objects detected by the algorithm. That step of manual analysis removed $\sim$86\% of the galaxies that were detected by the algorithm. Figure~\ref{false_positives} shows examples of objects that were detected by the algorithm as peculiar, but are not peculiar objects by manual inspection. As the figure shows, these objects include objects that are not rare and can be considered false positives, as well as objects that their peculiarity is not clear of is not of astronomical origin. Because the ``peculiarity'' of a galaxy is not strictly defined, it is possible that some objects of interest were rejected, but the prevalence of such objects is expected to be low.

\begin{figure}
\centering
\includegraphics[scale=0.6]{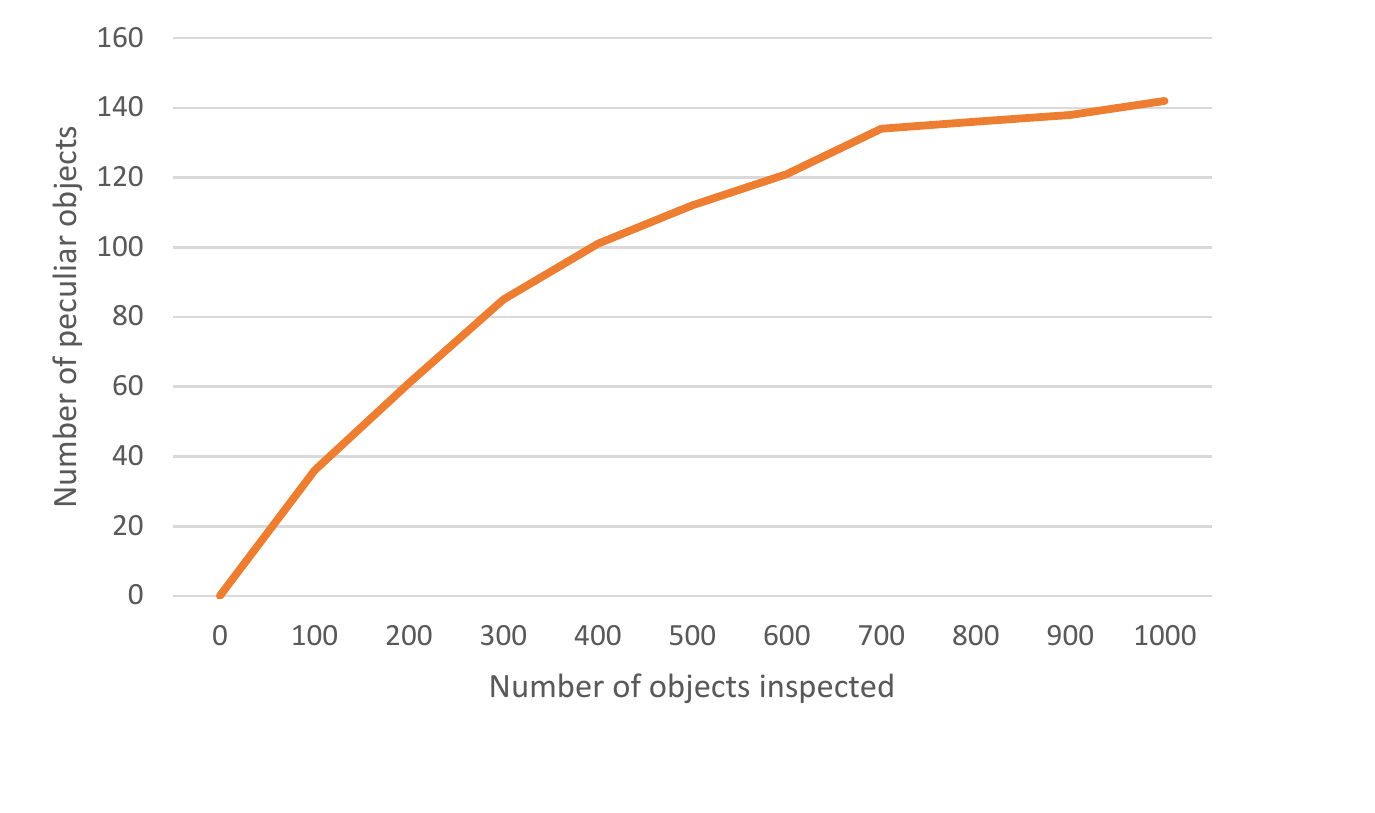}
\caption{The frequency of the number of objects detected manually among the objects detected by the algorithm.}
\label{frequency}
\end{figure}

The algorithm reduces the data by selecting a subset in which the frequency of outlier galaxy images is far higher than in the entire dataset, making the manual analysis practical also for larger datasets. Figure~\ref{frequency} shows the number of objects detected manually among the objects detected automatically, ranked by their distance as described in Section~\ref{method}. Naturally, the number of detected objects increases when the number of objects being inspected manually gets larger. But the graph also shows that the frequency of detected objects is higher among the galaxies with lower rank, therefore making it practical to perform manual analysis of the results.

\begin{figure}
\centering
\includegraphics[scale=0.4]{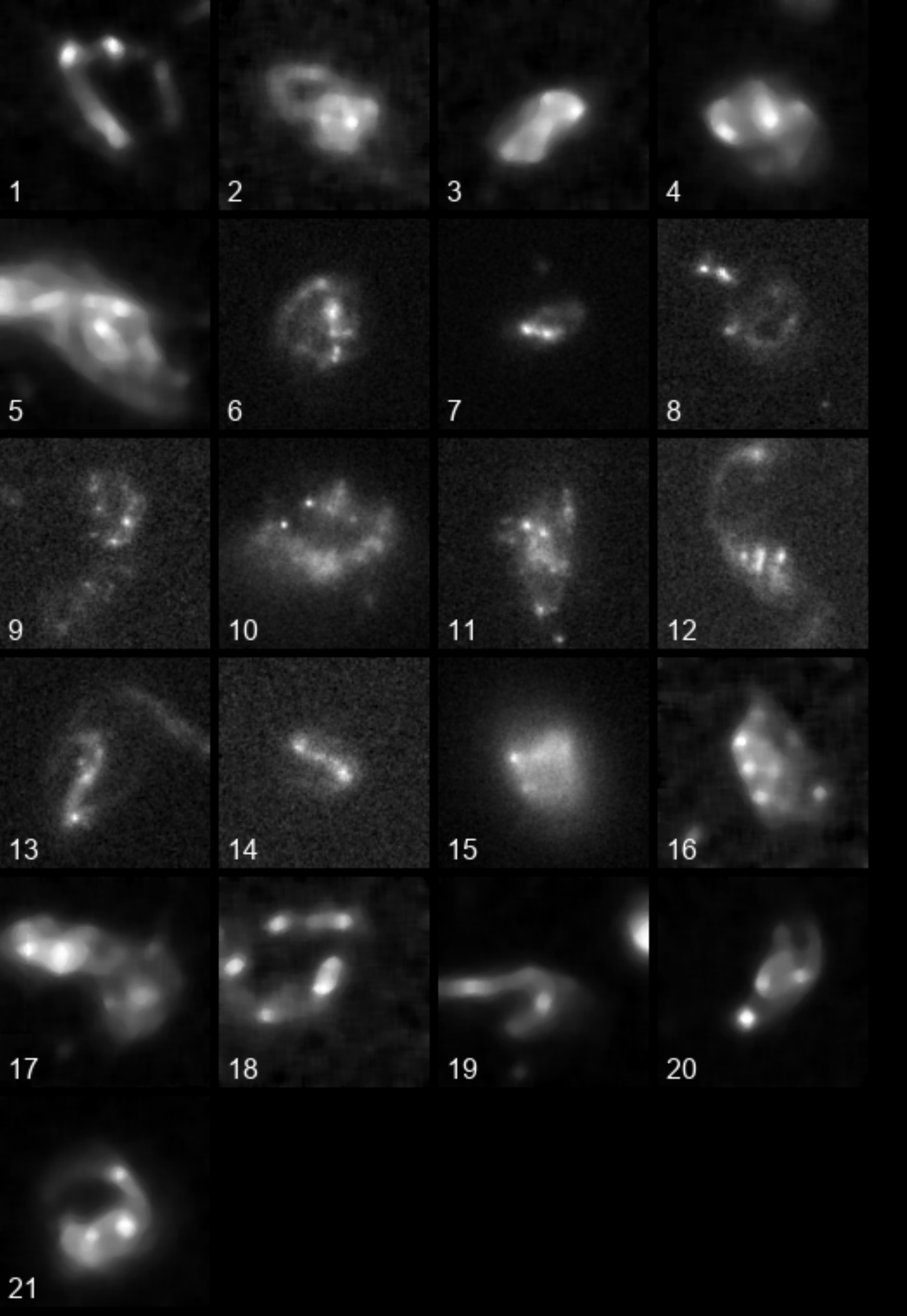}
\caption{Images of the detected objects listed in Table~\ref{embedded}.}
\label{embedded_images}
\end{figure}

\begin{figure}
\centering
\includegraphics[scale=0.4]{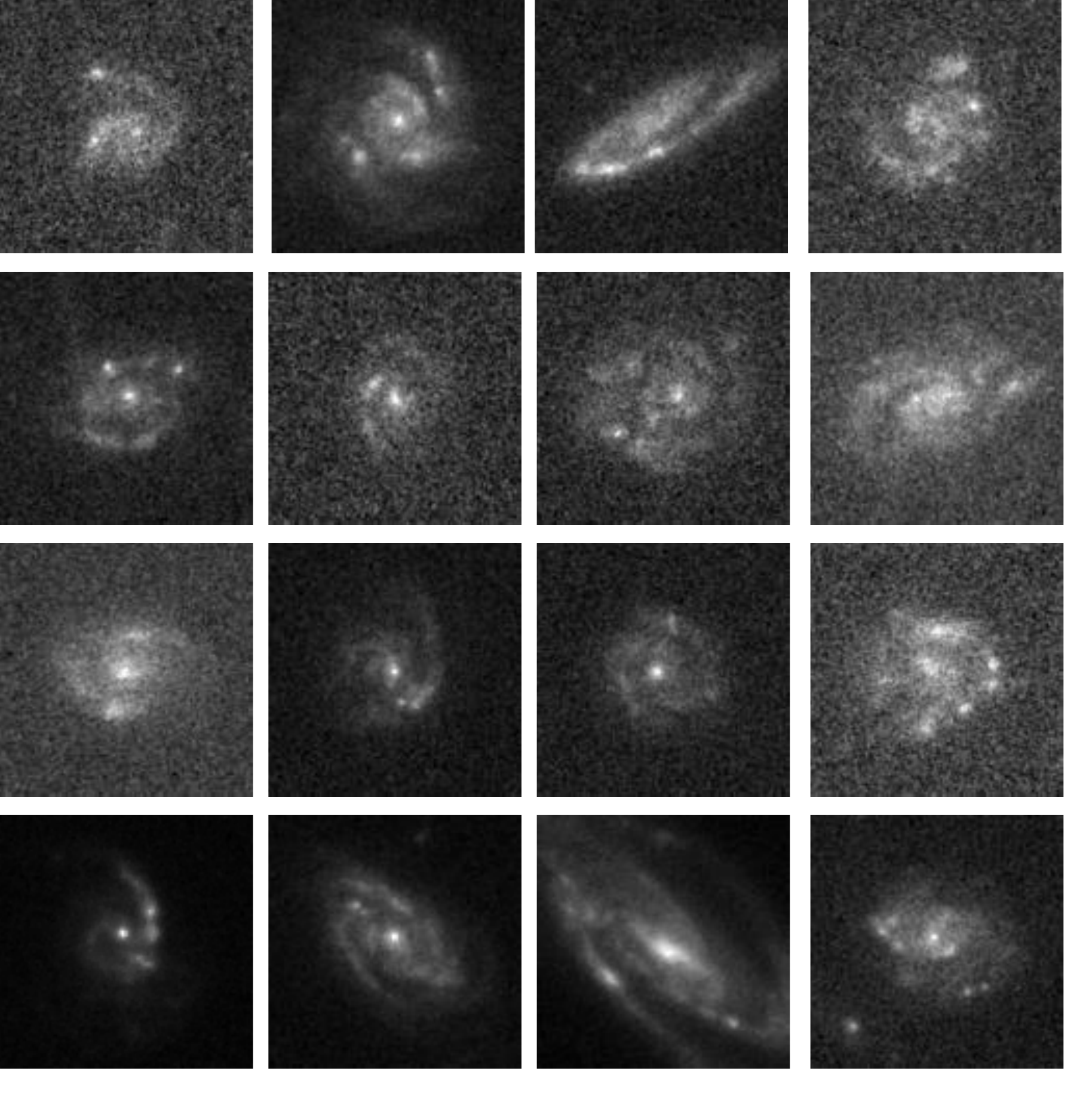}
\caption{Galaxies with clumps of stars that are not of apparent unusual morphology. These galaxies are very common in the HST sample, and were not detected by the algorithm.}
\label{clumps}
\end{figure}

\begin{figure}
\centering
\includegraphics[scale=0.4]{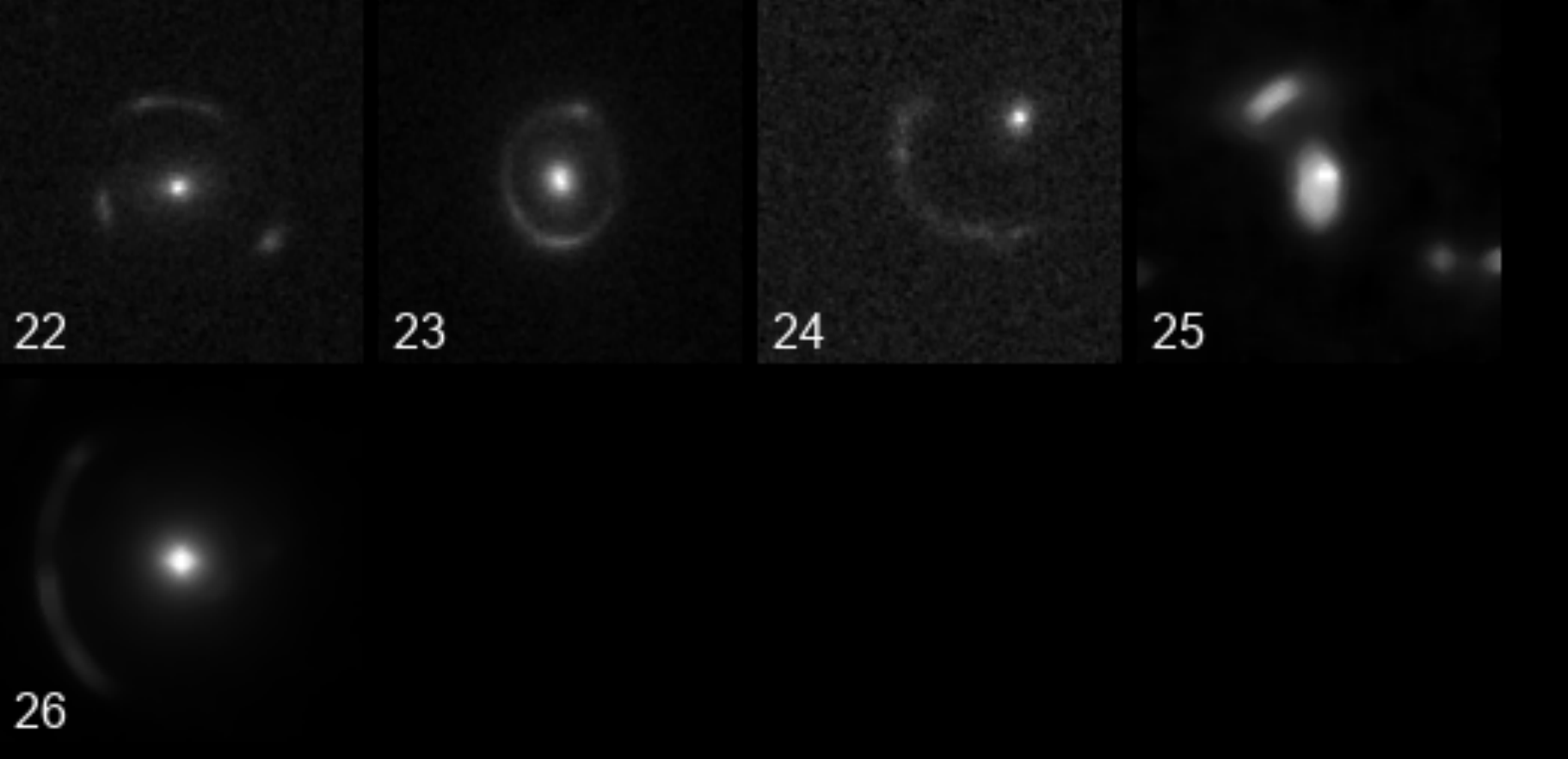}
\caption{Images of the objects suspected as gravitational lenses listed in Table~\ref{gravitational_lenses}.}
\label{gravitational_lenses_images}
\end{figure}

\begin{figure}
\centering
\includegraphics[scale=0.4]{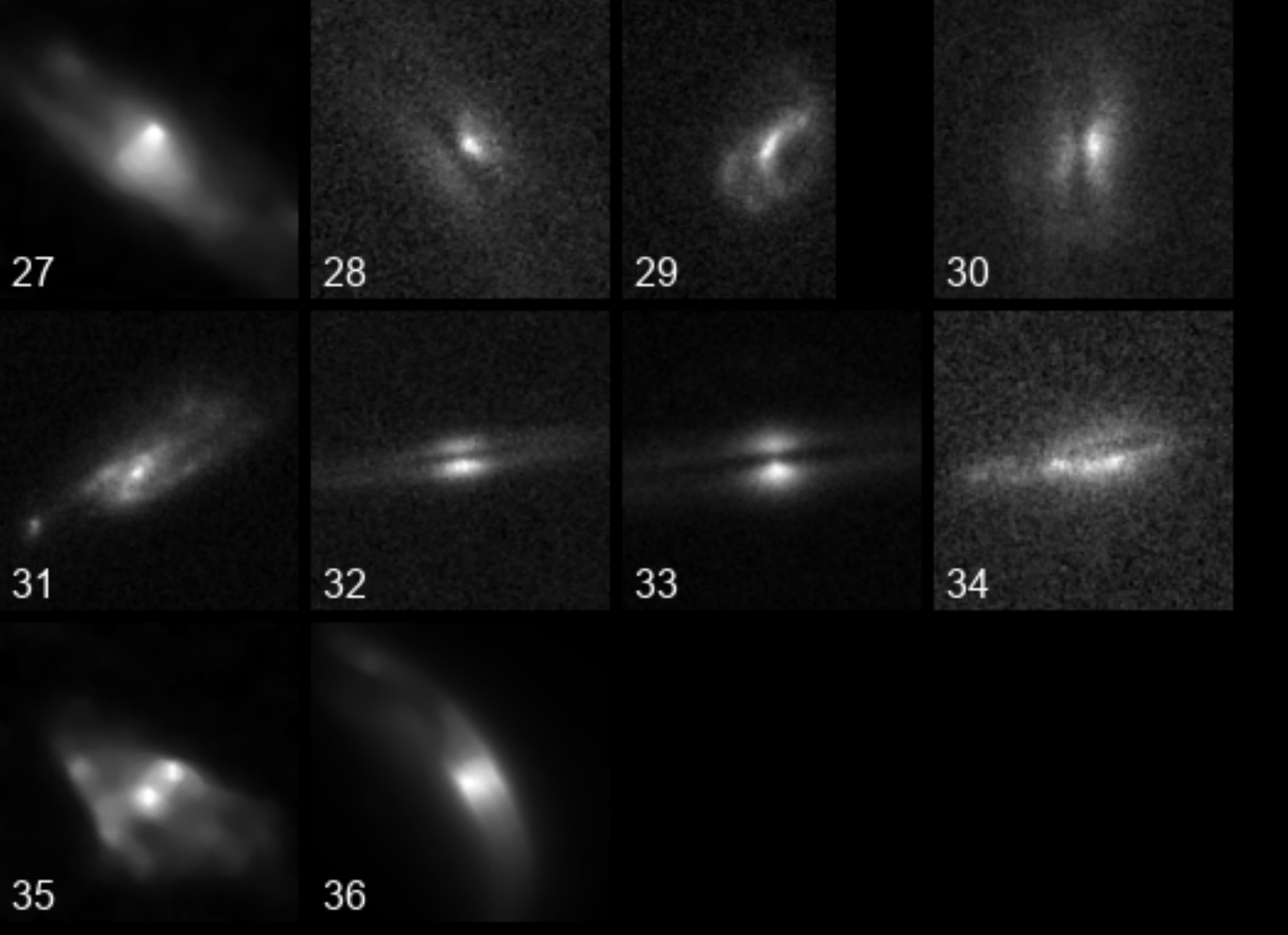}
\caption{Images of the detected edge-on galaxies with dust lanes listed in Table~\ref{edge_on_interacting}.}
\label{edge_on_interacting_images}
\end{figure}

\begin{figure}
\centering
\includegraphics[scale=0.4]{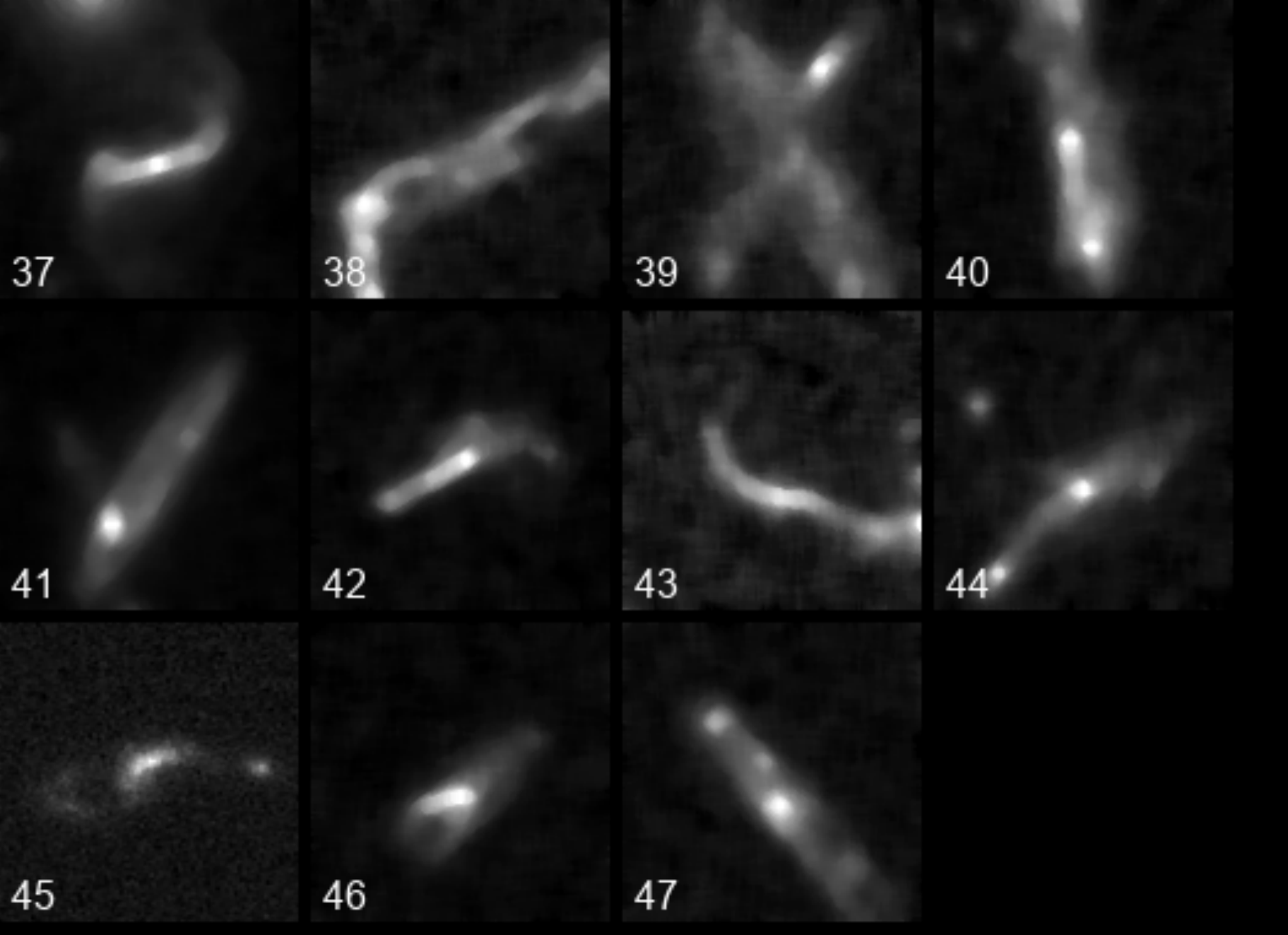}
\caption{Images of the detected objects with linear features listed in Table~\ref{linear}.}
\label{linear_images}
\end{figure}

\begin{figure}
\centering
\includegraphics[scale=0.4]{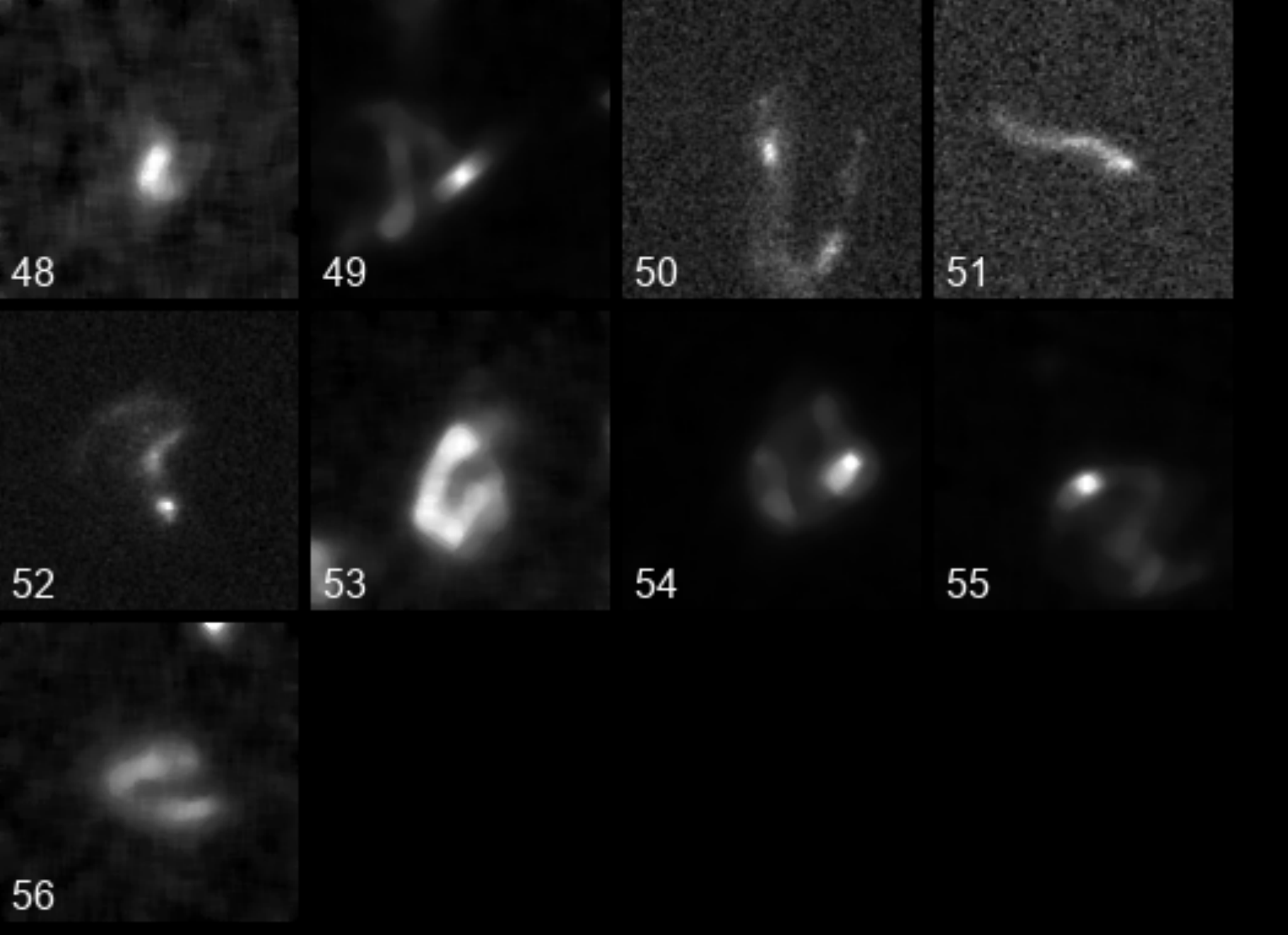}
\caption{Images of the detected objects listed in Table~\ref{one_arm}.}
\label{one_arm_images}
\end{figure}

\begin{figure}
\centering
\includegraphics[scale=0.40]{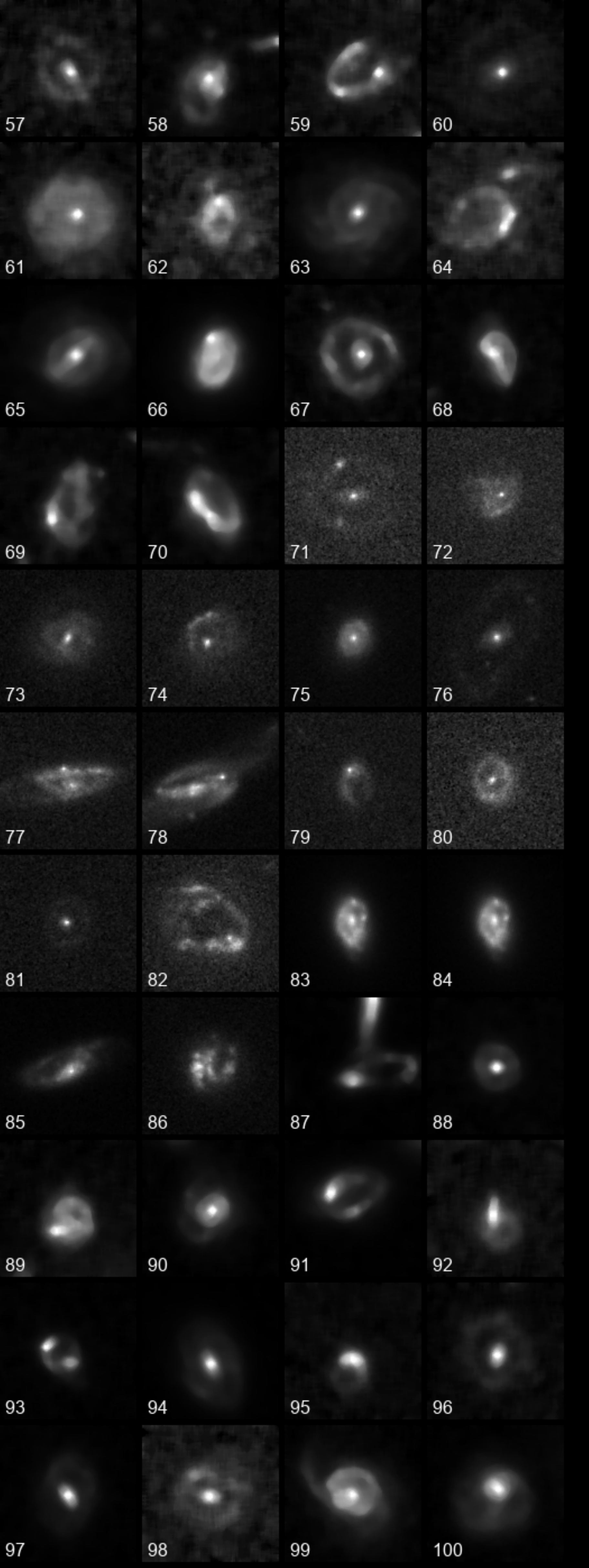}
\caption{Images of detected galaxies with ring features listed in Table~\ref{ring}.}
\label{ring_images}
\end{figure}

\begin{figure}
\centering
\includegraphics[scale=0.4]{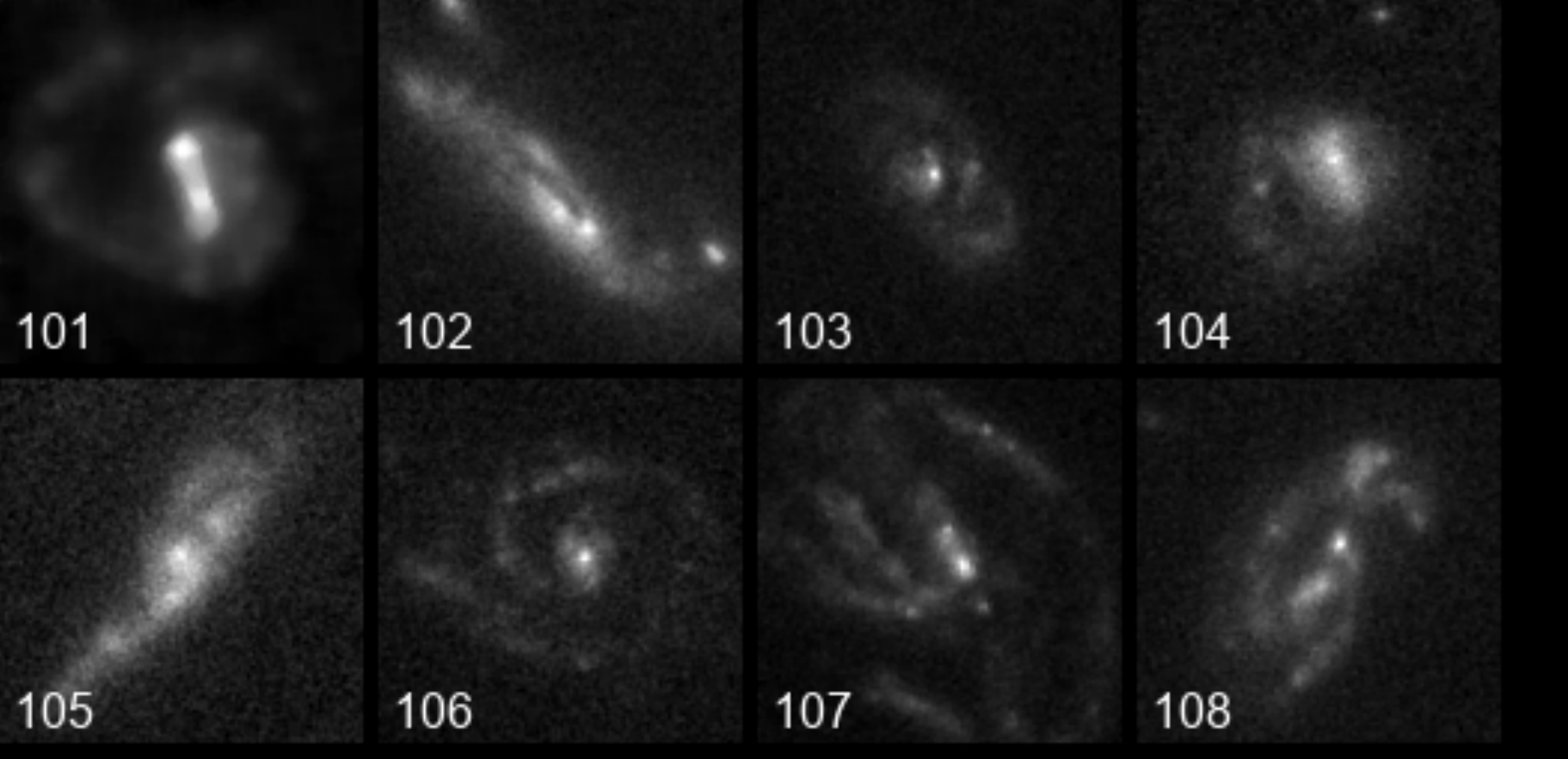}
\caption{Images of the detected spiral galaxies with detached segments listed in Table~\ref{detached}.}
\label{detached_images}
\end{figure}

\begin{figure}
\centering
\includegraphics[scale=0.40]{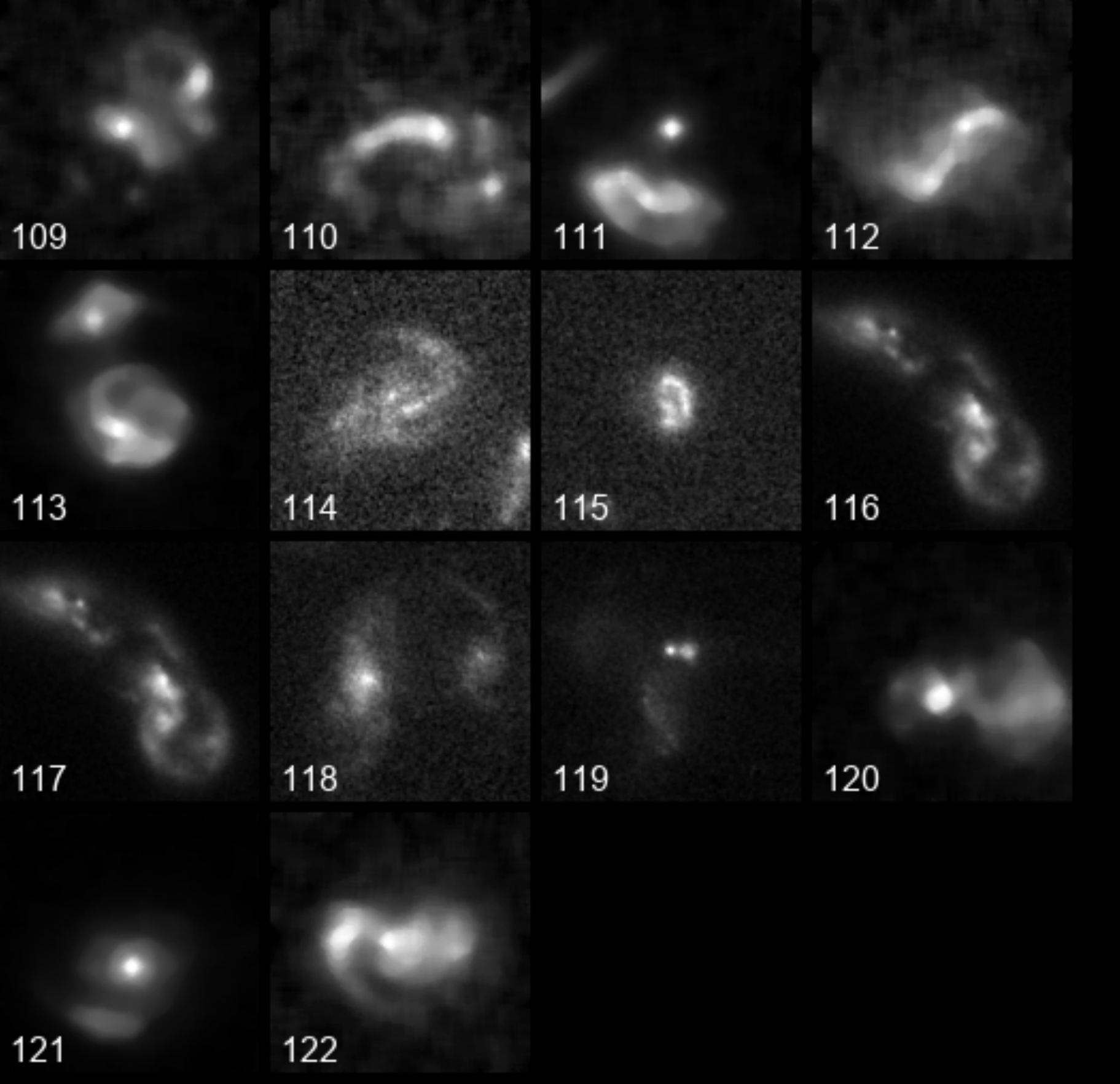}
\caption{Images of the tidally distorted object candidates listed in Table~\ref{tidally}.}
\label{tidally_images}
\end{figure}

\begin{figure}
\centering
\includegraphics[scale=0.4]{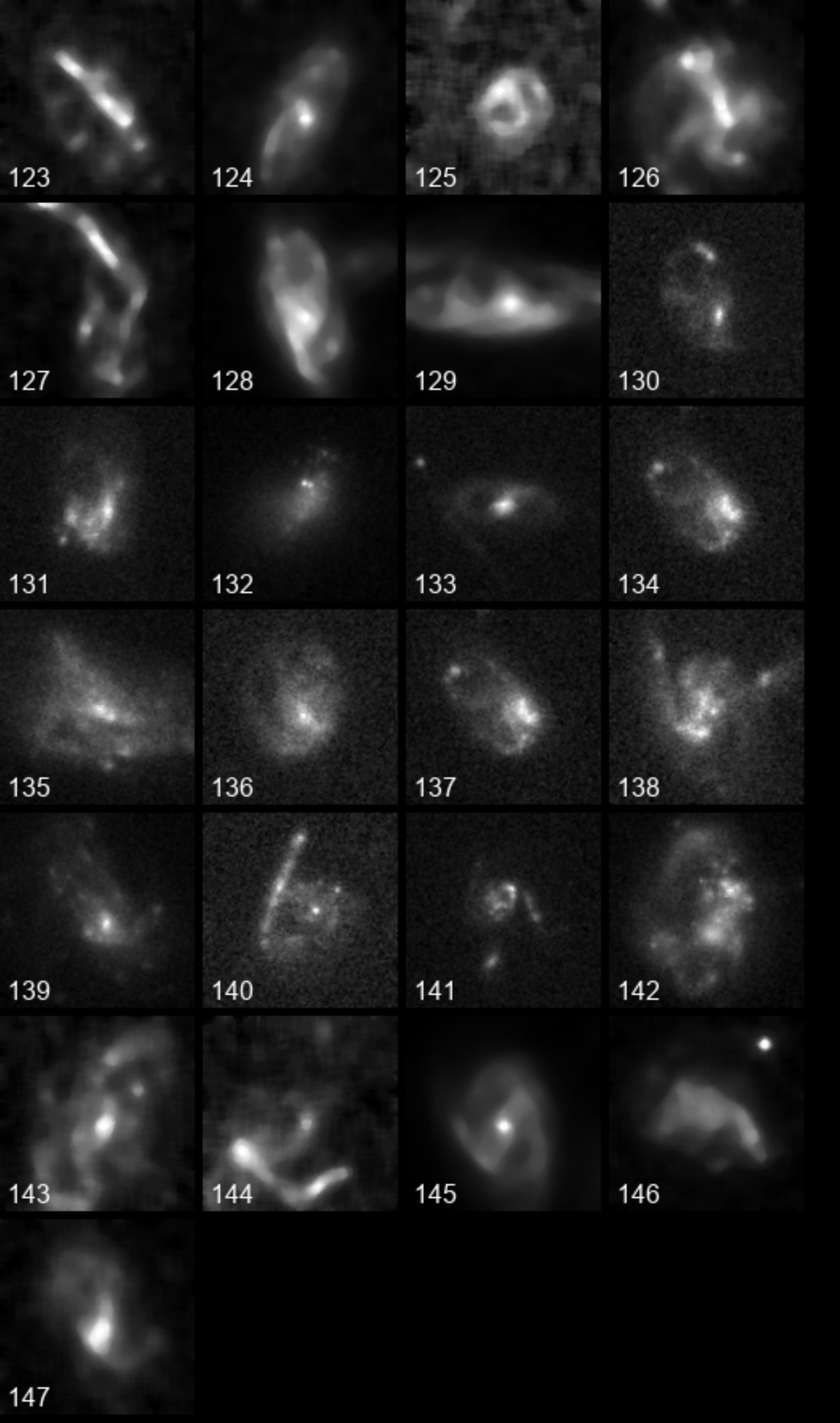}
\caption{Images of the other detected objects listed in Table~\ref{other}.}
\label{other_images}
\end{figure}

Tables~\ref{embedded} through~\ref{other} show the galaxies identified by the algorithm after removing manually $\sim$86\% of the objects that are not peculiar galaxies. Figures~\ref{embedded_images} through~\ref{other_images} show the corresponding images of the objects in the tables. The objects are separated into different types such as gravitational lenses, ring galaxies, objects with embedded point sources, interacting systems, objects with linear features, one-arm galaxies, galaxies with detached segments, tidally distorted interacting galaxies, and other galaxies.

Figure~\ref{edge_on_interacting_images} shows edge-on galaxies with dust lanes. These systems are not considered necessarily peculiar, but according to the results these forms are relatively rare in the HST sample. Figure~\ref{embedded_images} shows detected galaxies with embedded object. Giants clumps of stars embedded in galaxies are not rare in $0.5<z<3$ \citep{guo2015clumpy}. However, while many galaxies can have clumps of stars with no apparent unusual morphology as shown in Figure~\ref{clumps}, many of the galaxies in Figure~\ref{embedded_images} also have an unusual morphology that is different from the morphology of most galaxies with embedded bright clumps of stars. It is therefore possible that the morphology of these galaxies is the reason they are detected, rather than the clumps of stars embedded in them.


\begin{figure}
\centering
\includegraphics[scale=0.80]{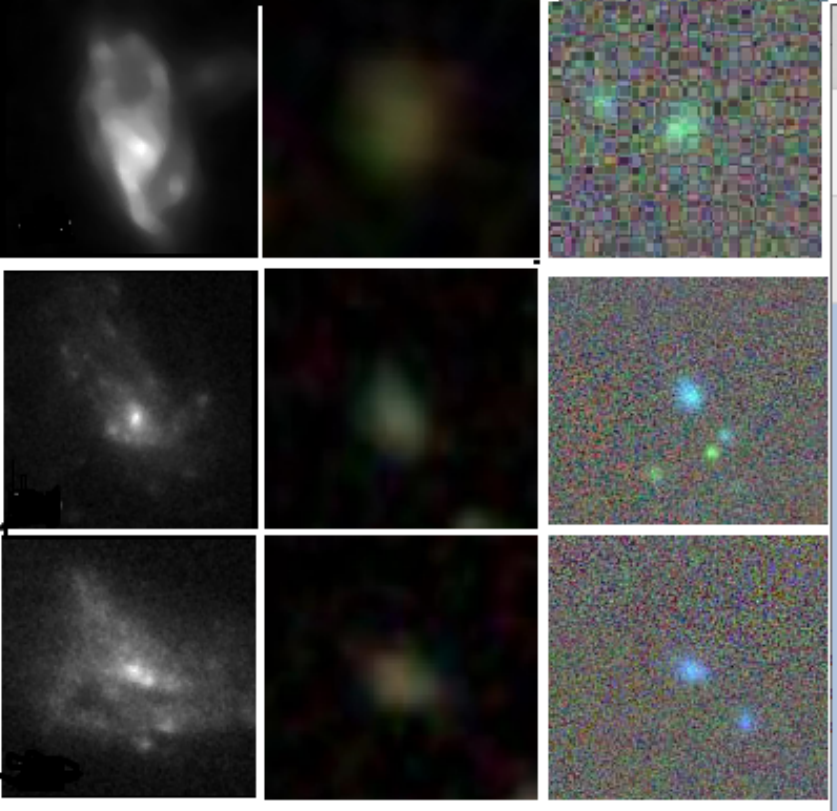}
\caption{Comparison of objects 128, 139 and 135 images in HST (left), SDSS (middle), and Pan-STARRS (right). As expected, the comparison shows that the Earth-based sky surveys do not provide sufficient details to identify the morphology of these galaxies.}
\label{comparison_sdss_ps}
\end{figure}

HST can provide details that cannot be obtained by Earth-based sky surveys, and therefore in many cases galaxies that seem visually peculiar in HST do not seem unusual when observed using Earth-based instruments. Figure~\ref{comparison_sdss_ps} shows several object in HST, SDSS and Pan-STARRS. As the comparison shows, SDSS and Pan-STARRS do not give sufficient details to identify the morphological features of these galaxies.

Table~\ref{gravitational_lenses} shows objects suspected as gravitational lenses. None of these suspected gravitational lenses are included in the CASTLES survey of gravitational lenses \citep{kochanek1999results}, the catalogue of gravitational lens candidates in SDSS \citep{inada2012sloan}, or a survey powered by a group finding algorithm \citep{wilson2016spectroscopic}. Two of the objects, 23 and 24, are included in the gravitational lens catalogue of \citep{faure2008first}.

\begin{table*}
{
\begin{tabular}{|l|c|c|c|c|c|c|c|c|c|c|}
\hline
ID & RA & Dec  & & ID & RA & Dec & & ID & RA & Dec \\
\hline
1 & 189.1622 & 62.1883 & & 2 & 215.1411 & 52.9442 & & 3 & 189.1906 & 62.2452 \\ 
4 & 188.9980 & 62.1668 & & 5 & 53.15489 & -27.857 & & 6 & 150.3146 & 1.68395 \\ 
7 & 150.1713 & 1.62978 & & 8 & 150.2772 & 1.91924 & & 9 & 150.0289 & 1.88902 \\ 
10 & 149.9010 & 1.85073 & & 11 & 150.0291 & 2.03546 & & 12 & 149.9216 & 2.20603 \\ 
13 & 150.0966 & 2.50137 & & 14 & 150.0506 & 2.47750 & & 15 & 150.7432 & 2.66317 \\ 
16 & 53.05503 & -27.699 & & 17 & 189.0773 & 62.2508 & & 18 & 53.06687 & -27.883 \\ 
19 & 189.1166 & 62.2854 & & 20 & 53.07838 & -27.878 & & 21 & 189.1230 & 62.1130 \\ 
\hline
\end{tabular}
\caption{The coordinates of detected objects with embedded point sources.}
\label{embedded}
}
\end{table*}


\begin{table*}
{
\begin{tabular}{|l|c|c|c|c|c|c|c|c|c|c|}
\hline
ID & RA & Dec  & & ID & RA & Dec & & ID & RA & Dec \\
\hline
22 & 149.8789 & 2.57436 & & 23 & 150.1594 & 2.69273 & & 24 & 150.0772 & 2.64584 \\ 
25 & 53.00104 & -27.770 & & 26 & 34.40478 & -5.2248 & &  & &  \\
\hline
\end{tabular}
\caption{Right ascension and declination (in degrees) of the galaxies suspected as gravitational lenses detected in the dataset.}
\label{gravitational_lenses}
}
\end{table*}

\begin{table*}
{
\begin{tabular}{|l|c|c|c|c|c|c|c|c|c|c|}
\hline
ID & RA & Dec  & & ID & RA & Dec & & ID & RA & Dec \\
\hline
27 & 215.3761 & 53.1241 & & 28 & 149.8313 & 1.59189 & & 29 & 150.0589 & 1.74697 \\ 
30 & 150.0610 & 1.64515 & & 31 & 150.3063 & 1.81053 & & 32 & 149.8813 & 1.88521 \\ 
33 & 149.8668 & 2.05173 & & 34 & 150.2041 & 2.80623 & & 35 & 189.0973 & 62.2924 \\ 
36 & 53.07250 & -27.822 & &  & &    & &  & &  \\
\hline
\end{tabular}
\caption{Celestial coordinates of objects that are possible edge-on galaxies with dust lanes.}
\label{edge_on_interacting}
}
\end{table*}

\begin{table*}
{
\begin{tabular}{|l|c|c|c|c|c|c|c|c|c|c|}
\hline
ID & RA & Dec  & & ID & RA & Dec & & ID & RA & Dec \\
\hline
37 & 215.2375 & 53.0477 & & 38 & 215.2534 & 53.0987 & & 39 & 215.1405 & 53.0041 \\ 
40 & 189.1852 & 62.1949 & & 41 & 189.2101 & 62.3432 & & 42 & 53.12404 & -27.878 \\ 
43 & 189.2438 & 62.1366 & & 44 & 189.2722 & 62.1792 & & 45 & 150.6507 & 1.64533 \\ 
46 & 189.0420 & 62.2196 & & 47 & 188.9525 & 62.1982 & &  & &  \\ 
\hline
\end{tabular}
\caption{Coordinates of detected objects with linear features.}
\label{linear}
}
\end{table*}

\begin{table*}
{
\begin{tabular}{|l|c|c|c|c|c|c|c|c|c|c|}
\hline
ID & RA & Dec  & & ID & RA & Dec & & ID & RA & Dec \\
\hline
48 & 214.8828 & 52.8360 & & 49 & 189.2587 & 62.3045 & & 50 & 150.6613 & 1.64342 \\ 
51 & 150.2551 & 1.88673 & & 52 & 150.1958 & 1.88558 & & 53 & 189.3390 & 62.1921 \\ 
54 & 189.0522 & 62.2440 & & 55 & 189.1127 & 62.2995 & & 56 & 34.31388 & -5.2024 \\ 
\hline
\end{tabular}
\caption{Coordinates of objects that are possible one-arm galaxies.}
\label{one_arm}
}
\end{table*}

\begin{table*}
{
\begin{tabular}{|l|c|c|c|c|c|c|c|c|c|c|}
\hline
ID & RA & Dec  & & ID & RA & Dec & & ID & RA & Dec \\
\hline
57 & 214.9668 & 52.8542 & & 58 & 189.1523 & 62.2768 & & 59 & 214.6585 & 52.7311 \\ 
60 & 214.6977 & 52.6933 & & 61 & 214.6236 & 52.7394 & & 62 & 214.6959 & 52.7275 \\ 
63 & 215.0946 & 52.9053 & & 64 & 214.9997 & 52.9886 & & 65 & 215.0564 & 53.0715 \\ 
66 & 215.3878 & 53.1364 & & 67 & 215.1373 & 53.0894 & & 68 & 53.12042 & -27.757 \\ 
69 & 189.2676 & 62.2110 & & 70 & 189.2785 & 62.1685 & & 71 & 150.1785 & 1.62206 \\ 
72 & 150.4450 & 1.72180 & & 73 & 150.3991 & 1.62907 & & 74 & 150.1361 & 1.67666 \\ 
75 & 149.7003 & 1.67250 & & 76 & 150.1600 & 1.92169 & & 77 & 150.1017 & 2.05326 \\ 
78 & 149.9335 & 2.04432 & & 79 & 149.8189 & 2.07964 & & 80 & 150.6226 & 2.24475 \\ 
81 & 149.6272 & 2.19739 & & 82 & 150.4789 & 2.40455 & & 83 & 150.2822 & 2.46019 \\ 
84 & 150.2822 & 2.46019 & & 85 & 149.8462 & 2.85215 & & 86 & 149.7706 & 2.80442 \\ 
87 & 189.3322 & 62.1755 & & 88 & 53.19657 & -27.863 & & 89 & 189.3906 & 62.2292 \\ 
90 & 53.22027 & -27.854 & & 91 & 53.05081 & -27.679 & & 92 & 189.1096 & 62.1963 \\ 
93 & 189.1269 & 62.2739 & & 94 & 189.1349 & 62.1262 & & 95 & 189.1359 & 62.1229 \\ 
96 & 53.01292 & -27.718 & & 97 & 34.32599 & -5.2154 & & 98 & 34.39667 & -5.2660 \\ 
99 & 34.26670 & -5.1327 & & 100 & 34.32900 & -5.1332 & &  & &  \\ 
\hline
\end{tabular}
\caption{Galaxies with ring features.}
\label{ring}
}
\end{table*}

\begin{table*}
{
\begin{tabular}{|l|c|c|c|c|c|c|c|c|c|c|}
\hline
ID & RA & Dec  & & ID & RA & Dec & & ID & RA & Dec \\
\hline
101 & 215.2793 & 53.1822 & & 102 & 149.9231 & 1.72376 & & 103 & 149.7927 & 1.62935 \\ 
104 & 149.6892 & 1.64355 & & 105 & 150.6343 & 1.81815 & & 106 & 150.4635 & 1.88330 \\ 
107 & 150.4138 & 1.84758 & & 108 & 150.5652 & 2.16613 & &  & &  \\ 
\hline
\end{tabular}
\caption{Spiral galaxies with detached segments.}
\label{detached}
}
\end{table*}

\begin{table*}
{
\begin{tabular}{|l|c|c|c|c|c|c|c|c|c|c|}
\hline
ID & RA & Dec  & & ID & RA & Dec & & ID & RA & Dec \\
\hline
109 & 53.10084 & -27.831 & & 110 & 215.2063 & 53.1576 & & 111 & 53.02481 & -27.751 \\ 
112 & 189.0039 & 62.2173 & & 113 & 189.2672 & 62.3234 & & 114 & 150.3090 & 1.91672 \\ 
115 & 150.6876 & 1.97088 & & 116 & 149.8947 & 2.20815 & & 117 & 149.8947 & 2.20815 \\ 
118 & 150.6845 & 2.54897 & & 119 & 149.9031 & 2.82170 & & 120 & 189.0672 & 62.2663 \\ 
121 & 189.1247 & 62.2343 & & 122 & 34.44746 & -5.2467 & &  & &  \\ 
\hline
\end{tabular}
\caption{Tidally distorted interacting pairs.}
\label{tidally}
}
\end{table*}

\begin{table*}
{
\begin{tabular}{|l|c|c|c|c|c|c|c|c|c|c|}
\hline
ID & RA & Dec  & & ID & RA & Dec & & ID & RA & Dec \\
\hline
123 & 215.0012 & 52.9636 & & 124 & 189.1776 & 62.3058 & & 125 & 215.0720 & 52.9071 \\ 
126 & 214.9257 & 52.9287 & & 127 & 215.2056 & 52.9864 & & 128 & 215.2513 & 53.1415 \\ 
129 & 53.11494 & -27.767 & & 130 & 150.7439 & 1.61616 & & 131 & 150.1504 & 1.59564 \\ 
132 & 149.8757 & 1.61034 & & 133 & 150.7142 & 1.75447 & & 134 & 149.8485 & 1.79248 \\ 
135 & 149.7127 & 1.77889 & & 136 & 149.9777 & 1.83432 & & 137 & 149.8485 & 1.79248 \\ 
138 & 150.4372 & 1.99945 & & 139 & 150.5846 & 2.19190 & & 140 & 149.7653 & 2.26561 \\ 
141 & 150.3849 & 2.40657 & & 142 & 150.2759 & 2.45195 & & 143 & 189.4095 & 62.2583 \\ 
144 & 189.4532 & 62.2233 & & 145 & 189.0520 & 62.1946 & & 146 & 34.36355 & -5.2133 \\ 
147 & 34.31276 & -5.1375 & &  & &    & &  & &  \\ 
\hline
\end{tabular}
\caption{Other galaxies.}
\label{other}
}
\end{table*}

\subsection{Differences between regular and outlier galaxy images}

Because the analysis is based on numerical image content descriptors, it allows to identify descriptors that can discriminate between regular galaxy images and the outlier images. To identify these descriptors, the image content descriptors of the 147 outlier images were compared to the descriptors of 745 random regular images. The comparison was dine using the Linear Discriminant Analysis (LDA) scores, which can identify the features that can discriminate between the two classes. Table~\ref{features} shows some of the numerical image content descriptors with the highest LDA scores, and the means and standard deviation of the regular and outlier galaxy images. The description of the specific descriptors is provided in \citep{shamir2008wndchrm,shamir2010impressionism,shamir2013automatic,schutter2015galaxy,shamir2016morphology}.

\begin{table}
{
\begin{tabular}{|l|c|c|c|c|c|c|c|c|c|c|}
\hline
Descriptor & Regular & Outlier \\
               &  mean   & mean   \\  
\hline
Edge area & 2796$\pm$171 & 14578$\pm$ 65  \\
Tamura coarseness & 9.59$\pm$0.08 &  2.43$\pm$0.11 \\
Fractal bin 15 &   4004$\pm$98      &    16.79$\pm$0.45     \\
Fractal bin 14 &   3895$\pm$96      &    16.13$\pm$0.43     \\
Fractal bin 11 &    3445$\pm$89    &    13.32$\pm$0.33     \\
Fractal bin 13 &  3783$\pm$94    &    15.46$\pm$0.41     \\
Fractal bin 18 &    4288$\pm$105    &     18.57$\pm$0.53  \\
Fractal bin 12 &   3569$\pm$91    &    14.06$\pm$0.36     \\
\hline
\end{tabular}
\caption{The image content descriptors with the highest LDA separation between regular and outlier images.}
\label{features}
}
\end{table}

As the table shows, descriptors such as edge area and Tamura texture coarseness exhibit significant differences between regular and outlier galaxy images. An interesting observation is the fractality, computed by using box counting as described in \citep{lynch1991analysis,shamir2009early}. The fractality is much lower among outlier galaxy images compared to the regular images. That indicates that regular galaxies have higher fractality, which drops in the case of outlier images.

\subsection{Impact of the value of R}

As discussed in Section~\ref{method}, the value of R is used to avoid the impact of rare objects that have similar objects in the dataset. When analyzing large datasets, even a rare object is expected to appear more than once in the dataset. Therefore, if two rare objects that are very different from all other objects are present in the dataset, it could be that each one of them will be a similar neighbor to the other object. When using the distance from the closest neighbor, the similarity between the two objects will assign each of the objects with a relatively short distance, and therefore these objects might not be detected as peculiar.

To show the impact of the value of R, a simple experiment was done such that the 44 galaxies shown in Figure~\ref{ring_images} were combined with galaxies 40 through 44 shown in Figure~\ref{linear_images}. In that dataset, the ring galaxies are the regular images. When running the algorithm when R is set to 1 and observing the top 10 outliers returned by the algorithm, only objects 40 and 43 are detected among the top five outliers. That does not change when setting the value of R to 2 or 3. But when the value of R is set to 4, all objects 40 through 44 are detected among the top 10 outliers.

\subsection{Sensitivity to redshift}

\begin{figure}
\centering
\includegraphics[scale=0.7]{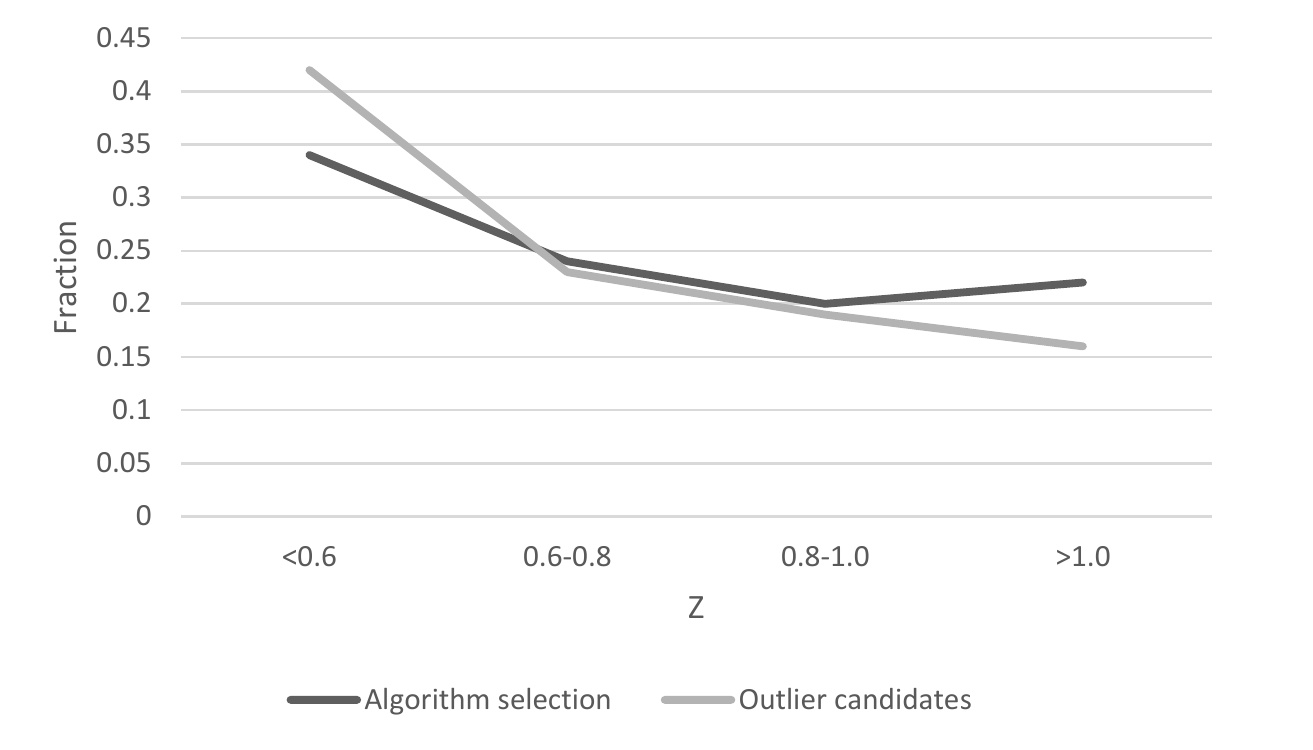}
\caption{The fraction of objects detected by the algorithm in different redshift ranges compared to the total number of objects detected by the algorithm, and the fraction of outlier candidates in each redshift range compared to the total number of outlier candidates determined after manual inspection. In the lower redshifts the fraction of outlier candidates is higher, while it is getting lower in the higher redshifts.}
\label{z_distribution}
\end{figure}

The ability of an algorithm to detect outlier galaxy is clearly a function of the redshift. Closer objects are generally brighter and can be observed with better details compared to distant object. It is expected that many objects with rare morphology at high redshift would not be identified as outliers by an algorithm or even by manual observation due to the small size and faint magnitude. Figure~\ref{z_distribution} shows the number of objects selected by the algorithm in each redshift range divided by the total number of objects detected by the algorithm. If also shows the number of objects determined as outlier candidates after manual inspection in each redshift range, divided by the total number of outlier candidates.

As the figure shows, the fraction of the objects selected after manual inspection is higher in the lower redshift ranges compared to the general population of objects selected by the algorithm, and lower in the higher redshift ranges of $z>1$. That distribution shows that a higher number of objects in the higher redshifts are detected by the algorithm but rejected after manual inspection, which indicates that in the higher redshifts the algorithm is less effective in identifying outlier galaxy candidates compared to the lower redshifts. That pattern can be expected given that galaxies at higher redshifts tend to be more difficult to inspect visually.

\section{Conclusions}
\label{conclusion}

Sky surveys can acquire substantial amounts of information that includes a very large number of galaxies. While it can be assumed that these databases contain rare objects of scientific interest, it is difficult to identify these objects among a large number of objects. Here an automatic method is applied to HST data, and identified several unusual extra-galactic objects. While the last step is manual, the algorithm reduces the data by two orders of magnitude, making the manual analysis practical. The objects identified by the algorithm can be used as target in future studies.

The catalogue is clearly incomplete, as the algorithm is not able to identify all rare objects of interest. For instance, just two gravitational lenses detected from the 67 detected gravitational lenses included in the catalogue of \citep{faure2008first}. However, since it is based on automation, it does not require substantial labor, and can therefore be applied in cases where the databases are far too large to allow manual analysis.

With the increasing importance of large-field surveys such as the ground-based Vera Rubin Observatory and the space-based Euclid, it is clear that manual analysis will not be sufficient to fully utilize the extreme imaging power of these instruments. While the efficacy of computer analysis cannot yet meet the accuracy level of manual analysis of an expert, computer analysis is required to approach these extremely large databases, and the ability to use the data acquired by current and future sky surveys is largely dependent on the availability and advancement of algorithm that can practically analyze these data.

\section*{Acknowledgment}

I would like to thank the anonymous reviewer for the insightful comments that helped to improve the manuscript. The research was funded by NSF grant AST-1903823.

\section*{Data availability}

The data underlying this article are available in the article. The research is based on observations made with the NASA/ESA Hubble Space Telescope, and obtained from the Hubble Legacy Archive, which is a collaboration between the Space Telescope Science Institute (STScI/NASA), the Space Telescope European Coordinating Facility (ST-ECF/ESA) and the Canadian Astronomy Data Centre (CADC/NRC/CSA).

\clearpage

\bibliographystyle{mnras}
\bibliography{hubble_peculiar}

\begin{thebibliography}{}
\makeatletter
\relax
\def\mn@urlcharsother{\let\do\@makeother \do\$\do\&\do\#\do\^\do\_\do\%\do\~}
\def\mn@doi{\begingroup\mn@urlcharsother \@ifnextchar [ {\mn@doi@}
  {\mn@doi@[]}}
\def\mn@doi@[#1]#2{\def\@tempa{#1}\ifx\@tempa\@empty \href
  {http://dx.doi.org/#2} {doi:#2}\else \href {http://dx.doi.org/#2} {#1}\fi
  \endgroup}
\def\mn@eprint#1#2{\mn@eprint@#1:#2::\@nil}
\def\mn@eprint@arXiv#1{\href {http://arxiv.org/abs/#1} {{\tt arXiv:#1}}}
\def\mn@eprint@dblp#1{\href {http://dblp.uni-trier.de/rec/bibtex/#1.xml}
  {dblp:#1}}
\def\mn@eprint@#1:#2:#3:#4\@nil{\def\@tempa {#1}\def\@tempb {#2}\def\@tempc
  {#3}\ifx \@tempc \@empty \let \@tempc \@tempb \let \@tempb \@tempa \fi \ifx
  \@tempb \@empty \def\@tempb {arXiv}\fi \@ifundefined
  {mn@eprint@\@tempb}{\@tempb:\@tempc}{\expandafter \expandafter \csname
  mn@eprint@\@tempb\endcsname \expandafter{\@tempc}}}

\bibitem[\protect\citeauthoryear{Abraham \& van~den Bergh}{Abraham \& van~den
  Bergh}{2001}]{abraham2001morphological}
Abraham R.~G.,  van~den Bergh S.,  2001, Science, 293, 1273

\bibitem[\protect\citeauthoryear{Abraham, Van Den~Bergh  \& Nair}{Abraham
  et~al.}{2003}]{abraham2003new}
Abraham R.~G.,  Van Den~Bergh S.,   Nair P.,  2003, Astrophysical Journal, 588,
  218

\bibitem[\protect\citeauthoryear{Amarbayasgalan, Jargalsaikhan  \&
  Ryu}{Amarbayasgalan et~al.}{2018}]{amarbayasgalan2018unsupervised}
Amarbayasgalan T.,  Jargalsaikhan B.,   Ryu K.~H.,  2018, Applied Sciences, 8,
  1468

\bibitem[\protect\citeauthoryear{Arp}{Arp}{1966}]{arp1966atlas}
Arp H.,  1966, Astrophysical Journal Supplement Series, 14, 1

\bibitem[\protect\citeauthoryear{Arp \& Madore}{Arp \&
  Madore}{1975}]{arp1975catalogue}
Arp H.~C.,  Madore B.~F.,  1975, The Observatory, 95, 212

\bibitem[\protect\citeauthoryear{Banerji et~al.,}{Banerji
  et~al.}{2010}]{banerji2010}
Banerji M.,  et~al., 2010, Monthly Notices of the Royal Astronomical Society,
  406, 342

\bibitem[\protect\citeauthoryear{Berriman et~al.,}{Berriman
  et~al.}{2004}]{berriman2004montage}
Berriman G.,  et~al., 2004, in Astronomical Data Analysis Software and Systems.
  p.~593

\bibitem[\protect\citeauthoryear{Bertin \& Arnouts}{Bertin \&
  Arnouts}{1996}]{bertin1996sextractor}
Bertin E.,  Arnouts S.,  1996, Astronomy and Astrophysics, 117, 393

\bibitem[\protect\citeauthoryear{Bettoni, Galletta, Garc{\'\i}a-Burillo  \&
  Rodr{\'\i}guez-Franco}{Bettoni et~al.}{2001}]{bettoni2001gas}
Bettoni D.,  Galletta G.,  Garc{\'\i}a-Burillo S.,   Rodr{\'\i}guez-Franco A.,
  2001, Astronomy \& Astrophysics, 374, 421

\bibitem[\protect\citeauthoryear{Buta}{Buta}{2017}]{buta2017galactic}
Buta R.~J.,  2017, Monthly Notices of the Royal Astronomical Society, 471, 4027

\bibitem[\protect\citeauthoryear{Casasola, Bettoni  \& Galletta}{Casasola
  et~al.}{2004}]{casasola2004gas}
Casasola V.,  Bettoni D.,   Galletta G.,  2004, Astronomy \& Astrophysics, 422,
  941

\bibitem[\protect\citeauthoryear{Cecotti}{Cecotti}{2020}]{cecotti2020rotation}
Cecotti H.,  2020, International Journal of Machine Learning and Cybernetics,
  pp 1--15

\bibitem[\protect\citeauthoryear{Chen, Yeo, Lee, Lau  \& Jin}{Chen
  et~al.}{2018}]{chen2018evolutionary}
Chen Z.,  Yeo C.~K.,  Lee B.~S.,  Lau C.~T.,   Jin Y.,  2018, Neurocomputing,
  309, 192

\bibitem[\protect\citeauthoryear{Cheng et~al.,}{Cheng
  et~al.}{2020}]{cheng2020optimizing}
Cheng T.-Y.,  et~al., 2020, Monthly Notices of the Royal Astronomical Society,
  493, 4209

\bibitem[\protect\citeauthoryear{Conselice}{Conselice}{2003}]{con03}
Conselice C.~J.,  2003, Astrophysical Journal Supplement Series, 147, 1

\bibitem[\protect\citeauthoryear{Davis \& Hayes}{Davis \&
  Hayes}{2014}]{davis2014sparcfire}
Davis D.~R.,  Hayes W.~B.,  2014, Astrophysical Journal, 790, 87

\bibitem[\protect\citeauthoryear{Dieleman, Willett  \& Dambre}{Dieleman
  et~al.}{2015}]{dieleman2015rotation}
Dieleman S.,  Willett K.~W.,   Dambre J.,  2015, Monthly Notices of the Royal
  Astronomical Society, 450, 1441

\bibitem[\protect\citeauthoryear{Faure et~al.,}{Faure
  et~al.}{2008}]{faure2008first}
Faure C.,  et~al., 2008, Astrophysical Journal Supplement Series, 176, 19

\bibitem[\protect\citeauthoryear{Finkelman, Funes~SJ  \& Brosch}{Finkelman
  et~al.}{2012}]{finkelman2012polar}
Finkelman I.,  Funes~SJ J.~G.,   Brosch N.,  2012, Monthly Notices of the Royal
  Astronomical Society, 422, 2386

\bibitem[\protect\citeauthoryear{Fogel \& Sagi}{Fogel \&
  Sagi}{1989}]{fogel1989gabor}
Fogel I.,  Sagi D.,  1989, Biological Cybernetics, 61, 103

\bibitem[\protect\citeauthoryear{Gillman et~al.,}{Gillman
  et~al.}{2020}]{gillman2020peculiar}
Gillman S.,  et~al., 2020, Monthly Notices of the Royal Astronomical Society,
  492, 1492

\bibitem[\protect\citeauthoryear{Goddard \& Shamir}{Goddard \&
  Shamir}{2020}]{goddard2020}
Goddard H.,  Shamir L.,  2020, Astrophysical Journal Supplement Series, 251, 28

\bibitem[\protect\citeauthoryear{Graham}{Graham}{2019}]{graham2019galaxy}
Graham A.~W.,  2019, Monthly Notices of the Royal Astronomical Society, 487,
  4995

\bibitem[\protect\citeauthoryear{Grogin et~al.,}{Grogin
  et~al.}{2011}]{grogin2011candels}
Grogin N.~A.,  et~al., 2011, Astrophysical Journal Supplement Series, 197, 35

\bibitem[\protect\citeauthoryear{Guo et~al.,}{Guo et~al.}{2015}]{guo2015clumpy}
Guo Y.,  et~al., 2015, Astrophysical Journal, 800, 39

\bibitem[\protect\citeauthoryear{Hadjidemetriou, Grossberg  \&
  Nayar}{Hadjidemetriou et~al.}{2001}]{hadjidemetriou2001spatial}
Hadjidemetriou E.,  Grossberg M.~D.,   Nayar S.~K.,  2001, in Proceedings of
  the IEEE Computer Society Conference on Computer Vision and Pattern
  Recognition. pp~I--I

\bibitem[\protect\citeauthoryear{Haralick, Shanmugam  \& Dinstein}{Haralick
  et~al.}{1973}]{haralick1973textural}
Haralick R.~M.,  Shanmugam K.,   Dinstein I.~H.,  1973, IEEE Transactions on
  Systems, Man, and Cybernetics, pp 610--621

\bibitem[\protect\citeauthoryear{Hosny, Elaziz, Selim  \& Darwish}{Hosny
  et~al.}{2020}]{hosny2020classification}
Hosny K.,  Elaziz M.,  Selim I.,   Darwish M.,  2020, Astronomy and Computing,
  p. 100383

\bibitem[\protect\citeauthoryear{Huertas-Company et~al.,}{Huertas-Company
  et~al.}{2009}]{huertas2009robust}
Huertas-Company M.,  et~al., 2009, Astronomy and Astrophysics, 497, 743

\bibitem[\protect\citeauthoryear{Huertas-Company et~al.,}{Huertas-Company
  et~al.}{2015a}]{huertas2015catalog}
Huertas-Company M.,  et~al., 2015a, arXiv preprint arXiv:1509.05429

\bibitem[\protect\citeauthoryear{Huertas-Company et~al.,}{Huertas-Company
  et~al.}{2015b}]{huertas2015morphologies}
Huertas-Company M.,  et~al., 2015b, arXiv preprint arXiv:1506.03084

\bibitem[\protect\citeauthoryear{Inada et~al.,}{Inada
  et~al.}{2012}]{inada2012sloan}
Inada N.,  et~al., 2012, Astronomical Journal, 143, 119

\bibitem[\protect\citeauthoryear{Jacobs et~al.,}{Jacobs
  et~al.}{2019}]{jacobs2019finding}
Jacobs C.,  et~al., 2019, Monthly Notices of the Royal Astronomical Society,
  484, 5330

\bibitem[\protect\citeauthoryear{Kaviraj}{Kaviraj}{2010}]{kaviraj2010peculiar}
Kaviraj S.,  2010, Monthly Notices of the Royal Astronomical Society, 406, 382

\bibitem[\protect\citeauthoryear{Kochanek, Falco, Impey, Leh{\'a}r, McLeod  \&
  Rix}{Kochanek et~al.}{1999}]{kochanek1999results}
Kochanek C.,  Falco E.,  Impey C.,  Leh{\'a}r J.,  McLeod B.,   Rix H.-W.,
  1999, in AIP Conference Proceedings. pp 163--175

\bibitem[\protect\citeauthoryear{Koekemoer et~al.,}{Koekemoer
  et~al.}{2011}]{koekemoer2011candels}
Koekemoer A.~M.,  et~al., 2011, Astrophysical Journal Supplement Series, 197,
  36

\bibitem[\protect\citeauthoryear{Kuminski \& Shamir}{Kuminski \&
  Shamir}{2016}]{kuminski2016computer}
Kuminski E.,  Shamir L.,  2016, Astrophysical Journal Supplement Series, 223,
  20

\bibitem[\protect\citeauthoryear{Kuminski, George, Wallin  \& Shamir}{Kuminski
  et~al.}{2014}]{kum14}
Kuminski E.,  George J.,  Wallin J.,   Shamir L.,  2014, Publications of the
  Astronomical Society of the Pacific, 126, 959

\bibitem[\protect\citeauthoryear{Lim}{Lim}{1990}]{lim1990two}
Lim J.~S.,  1990, New Haven: Prentice Hall

\bibitem[\protect\citeauthoryear{Lintott et~al.,}{Lintott
  et~al.}{2009}]{lintott2009galaxy}
Lintott C.~J.,  et~al., 2009, Monthly Notices of the Royal Astronomical
  Society, 399, 129

\bibitem[\protect\citeauthoryear{Lynch, Hawkes  \& Buckland-Wright}{Lynch
  et~al.}{1991}]{lynch1991analysis}
Lynch J.,  Hawkes D.,   Buckland-Wright J.,  1991, Physics in Medicine \&
  Biology, 36, 709

\bibitem[\protect\citeauthoryear{Madore, Nelson  \& Petrillo}{Madore
  et~al.}{2009}]{madore2009atlas}
Madore B.~F.,  Nelson E.,   Petrillo K.,  2009, Astrophysical Journal
  Supplement Series, 181, 572

\bibitem[\protect\citeauthoryear{Margalef-Bentabol, Huertas-Company, Charnock,
  Margalef-Bentabol, Bernardi, Dubois, Storey-Fisher  \&
  Zanis}{Margalef-Bentabol et~al.}{2020}]{margalef2020detecting}
Margalef-Bentabol B.,  Huertas-Company M.,  Charnock T.,  Margalef-Bentabol C.,
   Bernardi M.,  Dubois Y.,  Storey-Fisher K.,   Zanis L.,  2020,
  arXiv:2003.08263

\bibitem[\protect\citeauthoryear{Margapuri, Shamir  \& Thapa}{Margapuri
  et~al.}{2020}]{venkat2020}
Margapuri V. S.~K.,  Shamir L.,   Thapa B.,  2020, in 29th International
  Conference on Software Engineering and Data Engineering.

\bibitem[\protect\citeauthoryear{Mittal, Soorya, Nagrath  \& Hemanth}{Mittal
  et~al.}{2019}]{mittal2019data}
Mittal A.,  Soorya A.,  Nagrath P.,   Hemanth D.~J.,  2019, Earth Science
  Informatics, pp 1--17

\bibitem[\protect\citeauthoryear{Nair \& Abraham}{Nair \&
  Abraham}{2010}]{nair2010catalog}
Nair P.~B.,  Abraham R.~G.,  2010, Astrophysical Journal Supplement Series,
  186, 427

\bibitem[\protect\citeauthoryear{Nairn \& Lahav}{Nairn \&
  Lahav}{1997}]{nairn1997peculiar}
Nairn A.,  Lahav O.,  1997, Monthly Notices of the Royal Astronomical Society,
  286, 969

\bibitem[\protect\citeauthoryear{Peng, Ho, Impey  \& Rix}{Peng
  et~al.}{2002}]{pen02}
Peng C.~Y.,  Ho L.~C.,  Impey C.~D.,   Rix H.-W.,  2002, Astronomical Journal,
  124, 266

\bibitem[\protect\citeauthoryear{Rubner, Tomasi  \& Guibas}{Rubner
  et~al.}{2000}]{rubner2000earth}
Rubner Y.,  Tomasi C.,   Guibas L.~J.,  2000, International Journal of Computer
  Vision, 40, 99

\bibitem[\protect\citeauthoryear{Ruzon \& Tomasi}{Ruzon \&
  Tomasi}{2001}]{ruzon2001edge}
Ruzon M.~A.,  Tomasi C.,  2001, IEEE Transactions on Pattern Analysis and
  Machine Intelligence, 23, 1281

\bibitem[\protect\citeauthoryear{Schutter \& Shamir}{Schutter \&
  Shamir}{2015}]{schutter2015galaxy}
Schutter A.,  Shamir L.,  2015, Astronomy and Computing, 12, 60

\bibitem[\protect\citeauthoryear{Scoville et~al.,}{Scoville
  et~al.}{2007}]{scoville2007cosmic}
Scoville N.,  et~al., 2007, Astrophysical Journal Supplement Series, 172, 1

\bibitem[\protect\citeauthoryear{Shamir}{Shamir}{2009}]{sha09}
Shamir L.,  2009, Monthly Notices of the Royal Astronomical Society, 399, 1367

\bibitem[\protect\citeauthoryear{Shamir}{Shamir}{2011}]{sha11}
Shamir L.,  2011, Astrophysical Journal, 736, 141

\bibitem[\protect\citeauthoryear{Shamir}{Shamir}{2012}]{shamir2012automatic}
Shamir L.,  2012, Journal of Computational Science, 3, 181

\bibitem[\protect\citeauthoryear{Shamir}{Shamir}{2016}]{shamir2016morphology}
Shamir L.,  2016, Publications of the Astronomical Society of the Pacific, 129,
  024003

\bibitem[\protect\citeauthoryear{Shamir}{Shamir}{2017}]{shamir2017udat}
Shamir L.,  2017, Astrophysics Source Code Library, p. ascl:1704.002

\bibitem[\protect\citeauthoryear{Shamir}{Shamir}{2020}]{shamir2020automatic}
Shamir L.,  2020, Monthly Notices of the Royal Astronomical Society, 491, 3767

\bibitem[\protect\citeauthoryear{Shamir \& Wallin}{Shamir \&
  Wallin}{2014}]{shamir2014automatic}
Shamir L.,  Wallin J.,  2014, Monthly Notices of the Royal Astronomical
  Society, 443, 3528

\bibitem[\protect\citeauthoryear{Shamir, Orlov, Eckley, Macura, Johnston  \&
  Goldberg}{Shamir et~al.}{2008}]{shamir2008wndchrm}
Shamir L.,  Orlov N.,  Eckley D.~M.,  Macura T.,  Johnston J.,   Goldberg
  I.~G.,  2008, Source Code for Biology and Medicine, 3, 13

\bibitem[\protect\citeauthoryear{Shamir, Ling, Scott, Hochberg, Ferrucci  \&
  Goldberg}{Shamir et~al.}{2009}]{shamir2009early}
Shamir L.,  Ling S.~M.,  Scott W.,  Hochberg M.,  Ferrucci L.,   Goldberg
  I.~G.,  2009, Osteoarthritis and Cartilage, 17, 1307

\bibitem[\protect\citeauthoryear{Shamir, Macura, Orlov, Eckley  \&
  Goldberg}{Shamir et~al.}{2010}]{shamir2010impressionism}
Shamir L.,  Macura T.,  Orlov N.,  Eckley D.~M.,   Goldberg I.~G.,  2010, ACM
  Transactions on Applied Perception, 7, 1

\bibitem[\protect\citeauthoryear{Shamir, Holincheck  \& Wallin}{Shamir
  et~al.}{2013}]{shamir2013automatic}
Shamir L.,  Holincheck A.,   Wallin J.,  2013, Astronomy and Computing, 2, 67

\bibitem[\protect\citeauthoryear{Simard}{Simard}{1999}]{sim99}
Simard L.,  1999, in Photometric Redshifts and the Detection of High Redshift
  Galaxies. p.~325

\bibitem[\protect\citeauthoryear{Tamura, Mori  \& Yamawaki}{Tamura
  et~al.}{1978}]{tamura1978textural}
Tamura H.,  Mori S.,   Yamawaki T.,  1978, IEEE Transactions on Systems, Man,
  and Cybernetics, 8, 460

\bibitem[\protect\citeauthoryear{Taylor, Jansen, Windhorst, Odewahn  \&
  Hibbard}{Taylor et~al.}{2005}]{taylor2005ubvr}
Taylor V.~A.,  Jansen R.~A.,  Windhorst R.~A.,  Odewahn S.~C.,   Hibbard J.~E.,
   2005, Astrophysical Journal, 630, 784

\bibitem[\protect\citeauthoryear{Teague}{Teague}{1980}]{teague1980image}
Teague M.~R.,  1980, Journal of the Optical Society of America, 70, 920

\bibitem[\protect\citeauthoryear{Timmis \& Shamir}{Timmis \&
  Shamir}{2017}]{timmis2017catalog}
Timmis I.,  Shamir L.,  2017, Astrophysical Journal Supplement Series, 231, 2

\bibitem[\protect\citeauthoryear{Wilson, Zabludoff, Ammons, Momcheva, Williams
  \& Keeton}{Wilson et~al.}{2016}]{wilson2016spectroscopic}
Wilson M.~L.,  Zabludoff A.~I.,  Ammons S.~M.,  Momcheva I.~G.,  Williams
  K.~A.,   Keeton C.~R.,  2016, Astrophysical Journal, 833, 194

\bibitem[\protect\citeauthoryear{Wu, Chen  \& Hsieh}{Wu
  et~al.}{1992}]{wu1992texture}
Wu C.-M.,  Chen Y.-C.,   Hsieh K.-S.,  1992, IEEE Transactions on Medical
  Imaging, 11, 141

\makeatother
\end{thebibliography}

\label{lastpage}

\end{document}